\documentclass[iop]{emulateapj}

\def\ra#1#2#3{#1$^{\rm h}$#2$^{\rm m}$#3$^{\rm s}$}
\def\dec#1#2#3{$#1^\circ#2'#3''$}
 
\def\lesssim{\mathrel{\hbox{\rlap{\hbox{\lower4pt\hbox{$\sim$}}}\hbox{$<$}}}}
\def\gtrsim{\mathrel{\hbox{\rlap{\hbox{\lower4pt\hbox{$\sim$}}}\hbox{$>$}}}}

\usepackage{epsfig,graphicx}
\usepackage{amsmath}
\usepackage{natbib}
\usepackage{comment}
\usepackage{url}
\usepackage{color}
\usepackage{gensymb}

\begin{document}

\title{Two New Calcium-Rich Gap Transients in Group and Cluster Environments}
\submitted{ApJ in press}
\email{rlunnan@astro.caltech.edu}

\def\cit{1}
\def\uw{2}
\def\okc{3}
\def\wis{4}
\def\cfa{5}
\def\lco{6}
\def\ucsb{7}
\def\car{8}
\def\ucb{9}
\def\stsci{10}
\def\sbu{11}
\def\ssc{12}
\def\ipac{13}
\def\lbnl{14}
\def\naoj{15}
\def\coo{16}

\author{R.~Lunnan\altaffilmark{\cit},
 M.~M.~Kasliwal\altaffilmark{\cit},
 Y.~Cao\altaffilmark{\cit,\uw},
 L.~Hangard\altaffilmark{\okc},
 O.~Yaron\altaffilmark{\wis},
 J.~T.~Parrent\altaffilmark{\cfa},
 C.~McCully\altaffilmark{\lco,\ucsb},
 A.~Gal-Yam\altaffilmark{\wis},
 J.~S.~Mulchaey\altaffilmark{\car},
 S.~Ben-Ami\altaffilmark{\cfa},
 A.~V.~Filippenko\altaffilmark{\ucb},
 C.~Fremling\altaffilmark{\okc},
 A.~S.~Fruchter\altaffilmark{\stsci},
 D.~A.~Howell\altaffilmark{\lco,\ucsb},
 J.~Koda\altaffilmark{\sbu},
 T.~Kupfer\altaffilmark{\cit},
 S.~R.~Kulkarni\altaffilmark{\cit},
 R.~Laher\altaffilmark{\ssc},
 F.~Masci\altaffilmark{\ipac},
 P.~E.~Nugent\altaffilmark{\ucb,\lbnl},
 E.~O.~Ofek\altaffilmark{\wis},
 M.~Yagi\altaffilmark{\naoj}, and
 Lin~Yan\altaffilmark{\ipac,\coo}.
}

\altaffiltext{\cit}{Department of Astronomy, California Institute of Technology, 1200 East California Boulevard, Pasadena, CA 91125, USA}
\altaffiltext{\uw}{eScience Institute and Astronomy Department, University of Washington, Seattle, WA 98195}
\altaffiltext{\okc}{Oskar Klein Centre, Physics Department, Stockholm University, SE-106 91 Stockholm, Sweden}
\altaffiltext{\wis}{Benoziyo Center for Astrophysics and the Helen Kimmel Center for Planetary Science, Weizmann Institute of Science, 76100 Rehovot, Israel}
\altaffiltext{\cfa}{Harvard-Smithsonian Center for Astrophysics, 60 Garden St, Cambridge, MA 02138, USA}
\altaffiltext{\lco}{Las Cumbres Observatory Global Telescope Network, 6740 Cortona Dr., Suite 102, Goleta, CA 93117, USA}
\altaffiltext{\ucsb}{Department of Physics, University of California, Santa Barbara, Broida Hall, Mail Code 9530, Santa Barbara, CA 93106-9530, USA}
\altaffiltext{\car}{The Observatories of the Carnegie Institution for Science, Pasadena, CA 91101, USA}
\altaffiltext{\ucb}{Department of Astronomy, University of California, Berkeley, CA 94720-3411, USA}
\altaffiltext{\stsci}{Space Telescope Science Institute, 3700 San Martin Drive, Baltimore, MD 21218, USA}
\altaffiltext{\sbu}{Department of Physics and Astronomy, Stony Brook University, Stony Brook, NY 11794-3800, USA}
\altaffiltext{\ssc}{Spitzer Science Center, California Institute of Technology, MS 314-6, Pasadena, CA 91125, USA}
\altaffiltext{\ipac}{Infrared Processing and Analysis Center, California Institute of Technology, MS 100-22, Pasadena, CA 91125, USA}
\altaffiltext{\lbnl}{Lawrence Berkeley National Laboratory, 1 Cyclotron Road, MS 50B-4206, Berkeley, CA 94720, USA}
\altaffiltext{\naoj}{Optical and Infrared Astronomy Division, National Astronomical Observatory of Japan, 2-21-1 Osawa, Mitaka, Tokyo, 181-8588, Japan}
\altaffiltext{\coo}{Caltech Optical Observatories, California Institute of Technology, 1200 East California Boulevard, Pasadena, CA 91125, USA}

\begin{abstract}
We present the Palomar Transient Factory discoveries and the photometric and spectroscopic observations of PTF11kmb and PTF12bho. We show that both transients have properties consistent with the class of calcium-rich gap transients, specifically lower peak luminosities and rapid evolution compared to ordinary supernovae, and a nebular spectrum dominated by [\ion{Ca}{2}] emission. A striking feature of both transients is their host environments: PTF12bho is an intra-cluster transient in the Coma Cluster, while PTF11kmb is located in a loose galaxy group, at a physical offset $\sim 150$~kpc from the most likely host galaxy. Deep Subaru imaging of PTF12bho rules out an underlying host system to a limit of $M_R > -8.0~{\rm mag}$, while \textit{Hubble Space Telescope} imaging of PTF11kmb reveals a marginal counterpart that, if real, could be either a background galaxy or a globular cluster. We show that the offset distribution of Ca-rich gap transients is significantly more extreme than that seen for Type Ia supernovae or even short-hard gamma-ray bursts (sGRBs). Thus, if the offsets are caused by a kick, they require larger kick velocities and/or longer merger times than sGRBs. We also show that almost all Ca-rich transients found to date are in group and cluster environments with elliptical host galaxies, indicating a very old progenitor population; the remote locations could partially be explained by these environments having the largest fraction of stars in the intra-group/intra-cluster light following galaxy-galaxy interactions.
\end{abstract}


\section{Introduction}
In the last decade, large transient surveys have discovered a number of previously unknown classes of astrophysical transients, expanding the parameter space of known explosions, including several new classes in the former luminosity ``gap'' between supernovae (SNe) and novae \citep{kas12}. One such new class is the Ca-rich gap transients, with the peculiar transient SN\,2005E as a prototype \citep{pgm+10}. Kasliwal et al. (2012; hereafter K12)\nocite{kkg+12} proposed the following five properties to define the class: (1) intermediate luminosity (``gap'' transients), (2) faster photometric evolution (rise and decline) than normal SNe, (3) photospheric velocities comparable to those of SNe, (4) rapid evolution to the nebular phase, and (5) a nebular spectrum dominated by calcium emission. The rapid evolution indicates a class of transients with low ejecta masses ($\lesssim 0.5~{\rm M}_{\odot}$), whereas the nebular spectra suggest an ejecta composition dominated by calcium \citep{pgm+10}.

The sample of Ca-rich gap transients is still small. Following the discovery of the first such object, SN\,2005E \citep{pgm+10}, the Palomar Transient Factory \citep{lkd+09} presented three additional objects with similar properties: PTF09dav, PTF10iuv, and PTF11bij (\citealt{skn+11}; K12). SN\,2012hn, found by CRTS \citep{ddm+09} and followed up by PESSTO, has since been added \citep{vyt+14}. In addition, there are a number of archival candidates noted for their Ca-rich nebular spectra \citep{fcs+03,pgm+10,pgc+11,kmn+10}, from which K12 determined that SN\,2007ke fits the Ca-rich gap transient criteria. The remainder of these archival candidates (6 total: SN\,2000ds, SN\,2001co, SN\,2003H, SN\,2003dg, SN\,2003dr, and SN\,2005cz) are sometimes referred to as ``Ca-rich supernovae'' (in particular, see \citealt{fcs+03}, who identified the first four such objects), but lack sufficient data to determine whether they constitute the same physical phenomenon as the more strictly defined Ca-rich gap transients. In particular, photometry is necessary to establish the low peak luminosity and low ejecta mass, and a high ratio of Ca to O in the nebular spectrum can also be seen in peculiar core-collapse and Type Ia SNe depending on the phase of the spectrum, and is in itself not sufficient \citep{vyt+14,pzt+04,vpc+09}. For the purposes of this paper, we only consider the objects that satisfy both the photometric and spectroscopic criteria of K12, though we sometimes refer to the extended sample as candidates. 

The progenitor system of Ca-rich gap transients is not known, but their host environments provide strong clues: out of the six confirmed members so far, only PTF09dav has a spiral host galaxy, while the rest have elliptical or S0 host galaxies, implying a predominantly old progenitor population. Even more striking, many are found at substantial offsets from their presumed host galaxies, suggesting either that they originate in extremely faint systems (such as globular clusters; \citealt{yks+13}), or alternatively that the progenitor has traveled from its birth site. Increasingly deep limits at the transient locations have failed to detect an underlying host system, favoring the latter interpretation \citep{ljp+13, llc+14, llj+16}. 

While the environments and locations strongly hint at an old progenitor population and thus a binary system progenitor, there are currently several proposed models for what the progenitor system actually is. \citet{pgm+10} suggested SN\,2005E may be the result of He detonation from accretion from a He white dwarf onto a CO white dwarf, one of several variants of WD-WD progenitors proposed \citep{wsl+11,sfk+12,dh15}. Another possibility is the tidal detonation of a He white dwarf by a neutron star or stellar-mass black hole \citep{met12,smk+15}. A better understanding of their origin is interesting not only from the point of view of binary evolution, but also from their nontrivial contribution to the Ca abundance: \citet{mkk14} showed that including the contribution from Ca-rich gap transients could resolve the discrepancy between the measured Ca abundances in galaxy clusters, and the abundances expected from core-collapse and Type Ia SN yields. As the sample of Ca-rich transients is still small, adding to the number of well-studied objects is necessary. Here, we present the discovery of two new members of the class of Ca-rich gap transients discovered by the Palomar Transient Factory (PTF11kmb and PTF12bho), as well as analysis of both the transients themselves and their host environments.

This paper is organized as follows. We present our observations of PTF11kmb and PTF12bho and their host environments in Section~\ref{sec:obs}. The transient properties are given in Section~\ref{sec:gap}, demonstrating that they are members of the class of Ca-rich gap transients. We discuss the host-galaxy properties in Section~\ref{sec:gal}. The implications for the origins of Ca-rich gap transients and a summary are provided in Section~\ref{sec:conc}. We present data on an additional Ca-rich gap transient candidate, PTF10hcw, in Appendix~\ref{sec:10hcw}. All calculations in this paper assume a $\Lambda$CDM cosmology with $H_0 = 70$~km~s$^{-1}$~Mpc$^{-1}$, $\Omega_{\rm M} = 0.27$ and $\Omega_{\Lambda} = 0.73$ \citep{ksd+11}. 

\section{Observations}
\label{sec:obs}

\subsection{Palomar Transient Factory Discoveries}

PTF11kmb and PTF12bho were found as part of the Palomar Transient Factory (PTF; \citealt{lkd+09, rkl+09}). PTF11kmb was discovered in data taken with the CFH12K 96-Megapixel camera \citep{rsv+08,ldr+10} mounted on the 48-inch Samuel Oschin Telescope at Palomar Observatory (P48) on 2011 Aug. 16.25 (UT dates are used throughout this paper) at J2000 coordinates $\alpha =$ \ra{22}{22}{53.61}, $\delta =$ \dec{+36}{17}{36.5}, and at a magnitude $r = 19.8$~mag. A spectrum was taken with the Low Resolution Imaging Spectrometer (LRIS; \citealt{occ+95}) on the 10-m Keck I telescope on 2011 Aug. 28, showing SN features consistent with a Type Ib SN (see \citealt{fil97} for a review of SN spectral classification) at a redshift $z = 0.017$ (Section~\ref{sec:spec}), and was reported as such in the discovery telegram \citep{gxb+11}; subsequent follow-up observations revealed it to be a Ca-rich gap transient (see also \citealt{fol15}). No host galaxy is seen underlying the SN position in the reference image. 

PTF12bho was discovered in P48 data on 2012 Feb. 25.25 at J2000 coordinates $\alpha =$ \ra{13}{01}{16.65}, $\delta =$ \dec{+28}{01}{18.5} and at a magnitude of $r=20.52$~mag. A spectrum taken with LRIS on 2012 Mar. 15 yields $z=0.023$ based on the SN features (Section~\ref{sec:spec}). Again, no underlying host galaxy is visible in the reference images; we note, however, that both the position and the redshift of PTF12bho are consistent with the Coma Cluster (Abell 1656).

\subsection{Light Curves}
\begin{figure}
\centering
\includegraphics[width=3.5in]{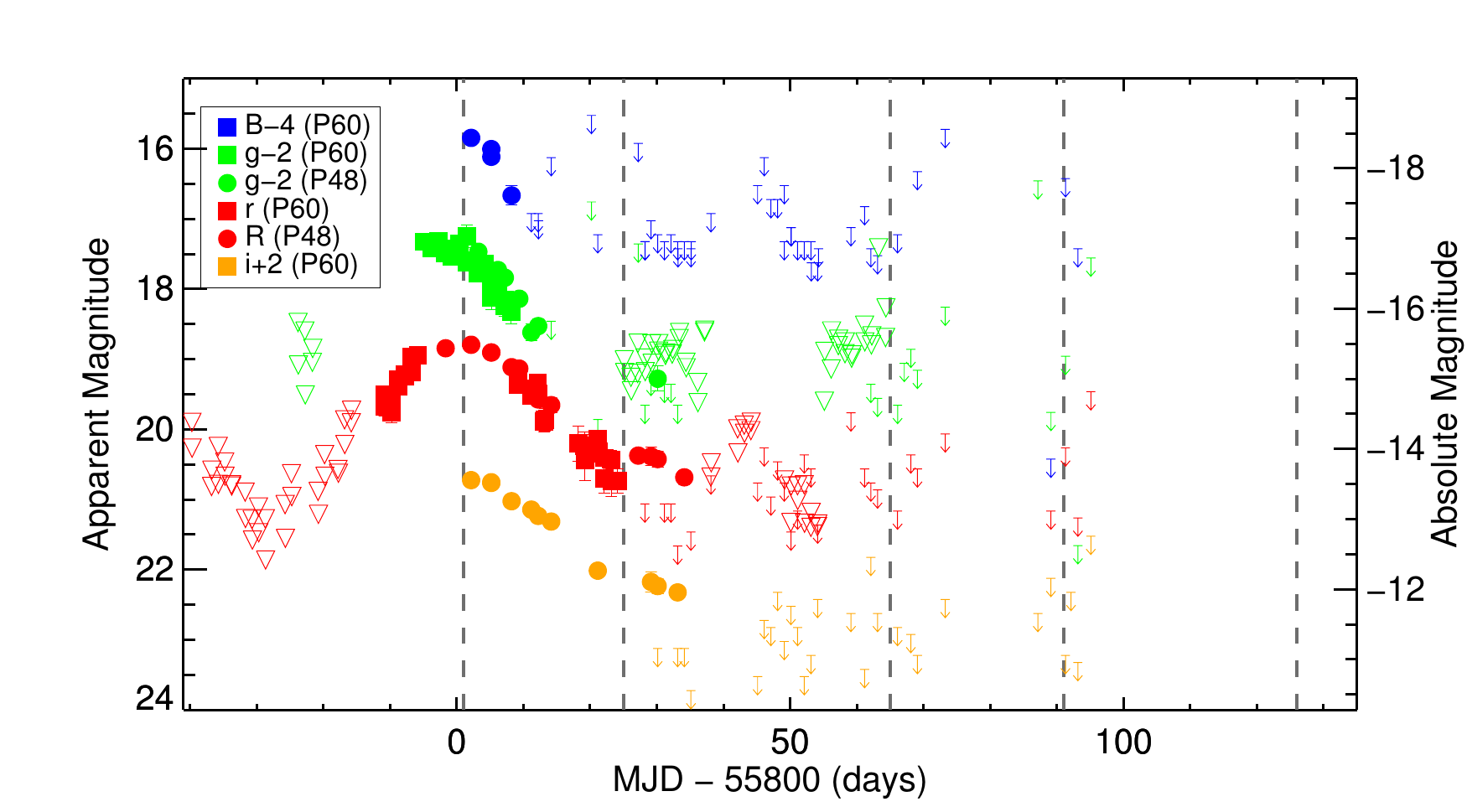}
\caption{Observed light curves of PTF11kmb. Different filters are offset for clarity as indicated in the legend. The dashed lines mark the spectroscopy epochs. Upper limits are 5$\sigma$ and are shown as triangles (P48) or arrows (P60). }
\label{fig:11kmb_lc}
\end{figure}

\begin{figure}
\centering
\includegraphics[width=3.5in]{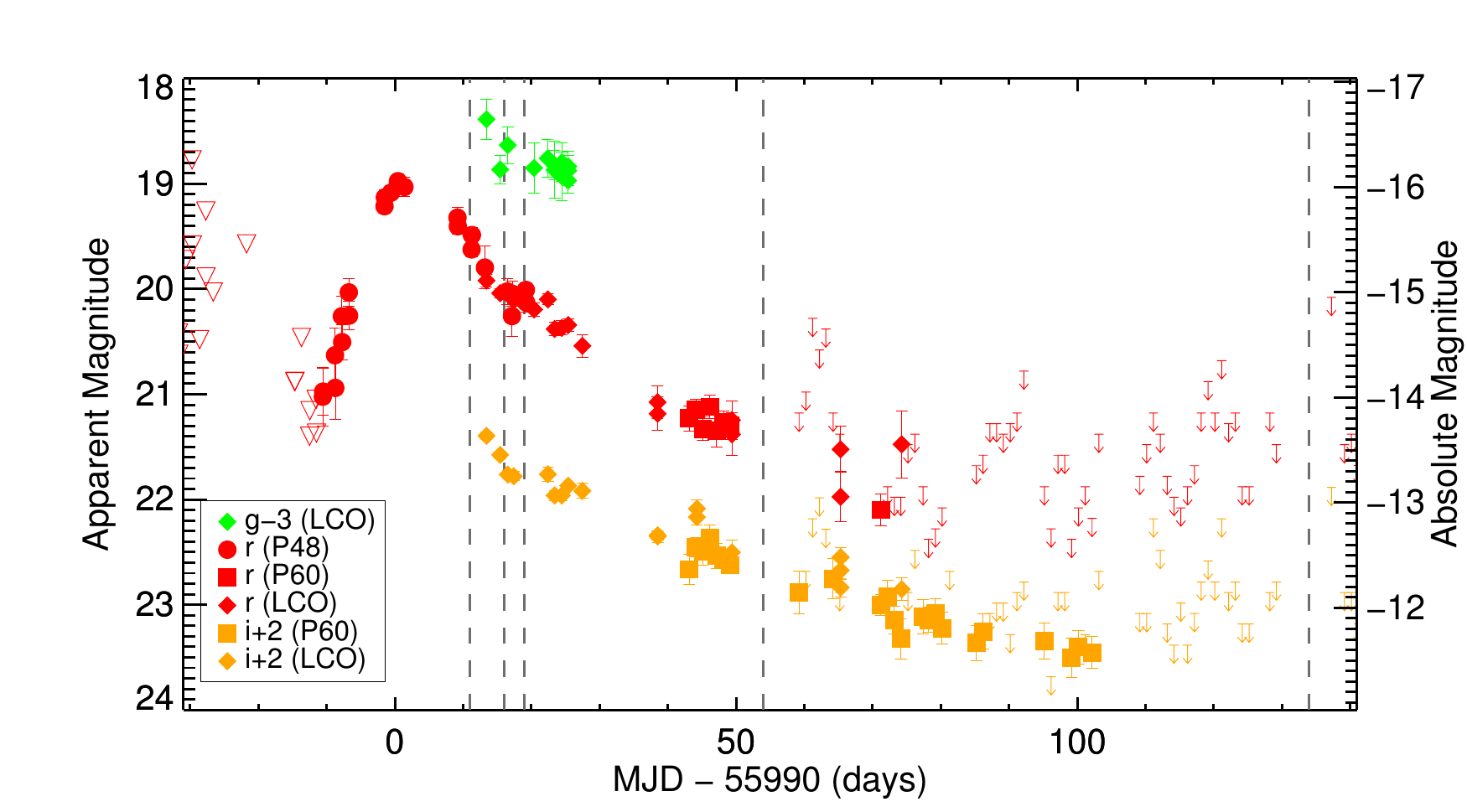}
\caption{Observed light curves of PTF12bho. Different filters are offset for clarity as indicated in the legend. The dashed lines mark the spectroscopy epochs. Upper limits are 5$\sigma$ and are shown as triangles (P48) or arrows (P60).}
\label{fig:12bho_lc}
\end{figure}

We obtained $R$- and $g$-band photometry of PTF11kmb and PTF12bho with the P48 CFH12K camera. Additional follow-up photometry was conducted with the automated 60-inch telescope at Palomar (P60; \citealt{cfm+06}) in the $Bgri$ bands, and with the Las Cumbres Observatory (LCO) Faulkes Telescope North in $gri$ \citep{bbb+13}. Host-subtracted point-spread function (PSF) photometry was obtained using the Palomar Transient Factory Image Differencing and Extraction (PTFIDE) pipeline \citep{mlr+17} on the P48 images, and the pipeline presented by \citet{fst+16} on the P60 images. LCO data were processed using the tools available at {\tt github.com/lcogt/banzai}, and PSF photometry was performed following \citet{vhs+16}. We correct the photometry for Galactic extinction following \citet{sf11}, with $E(B-V) = 0.092$ mag for PTF11kmb and $E(B-V) = 0.008$ mag for PTF12bho. All photometric data are listed in Table~\ref{tab:lc}, and shown in Figures~\ref{fig:11kmb_lc} and \ref{fig:12bho_lc}.

PTF12bho was also observed with the \textit{Swift} Ultra-Violet/Optical Telescope (UVOT; \citealt{uvot}) and the \textit{Swift} X-ray telescope (XRT; \citealt{xrt}) on 2012 March 17.8 for 3~ks. We process the \textit{Swift} data using the {\tt HEAsoft} package\footnote{\url{http://heasarc.nasa.gov/lheasoft/}} and find that PTF12bho is not detected with either instrument; the upper limit from UVOT is $u > 22.95~{\rm mag}$ (AB mag, 3$\sigma$). The XRT count limit is $< 4.6 \times 10^{-3}~{\rm counts~s}^{-1}$ (3$\sigma$); using the PIMMS tools\footnote{\url{http://heasarc.gsfc.nasa.gov/cgi-bin/Tools/w3pimms/w3pimms.pl}} we calculate a corresponding flux limit $F_X < 1.6\times 10^{-13}~{\rm erg~cm}^{-2}~{\rm s}^{-1}$ if assuming a power-law source with index $\Gamma = 2$ (0.3-10~keV).

\subsection{Spectroscopy}
\begin{figure*}
\centering
\begin{tabular}{cc}
\includegraphics[width=3.5in]{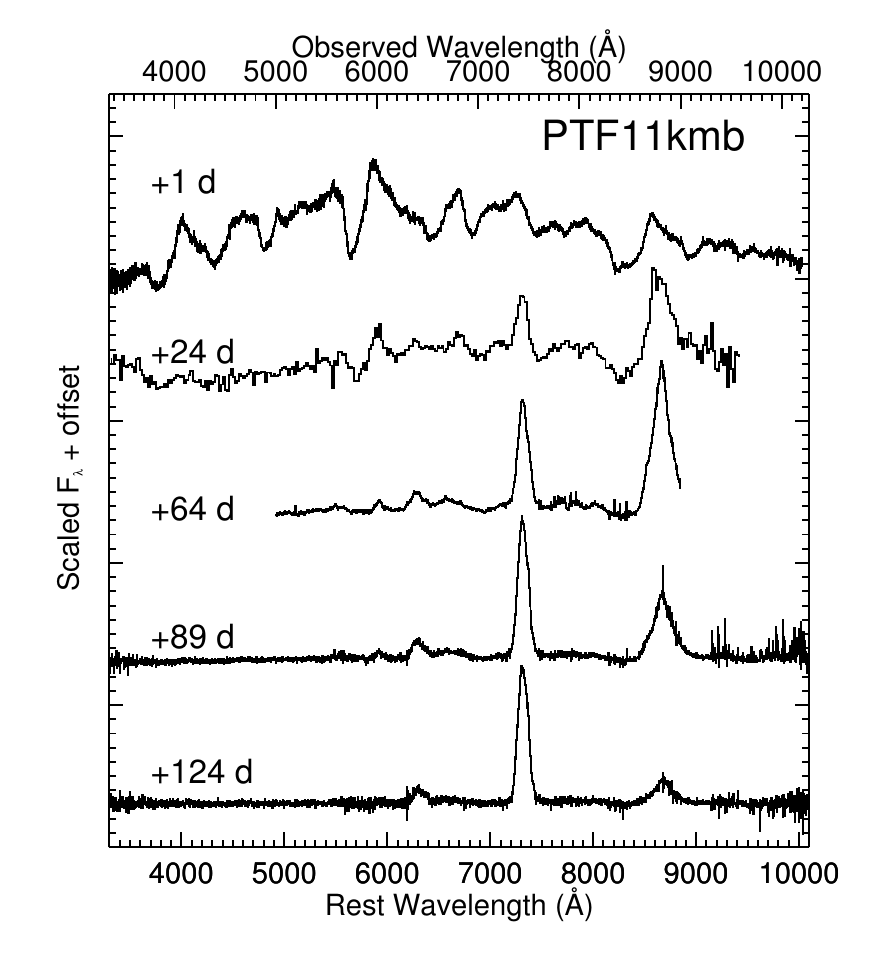} & \includegraphics[width=3.5in]{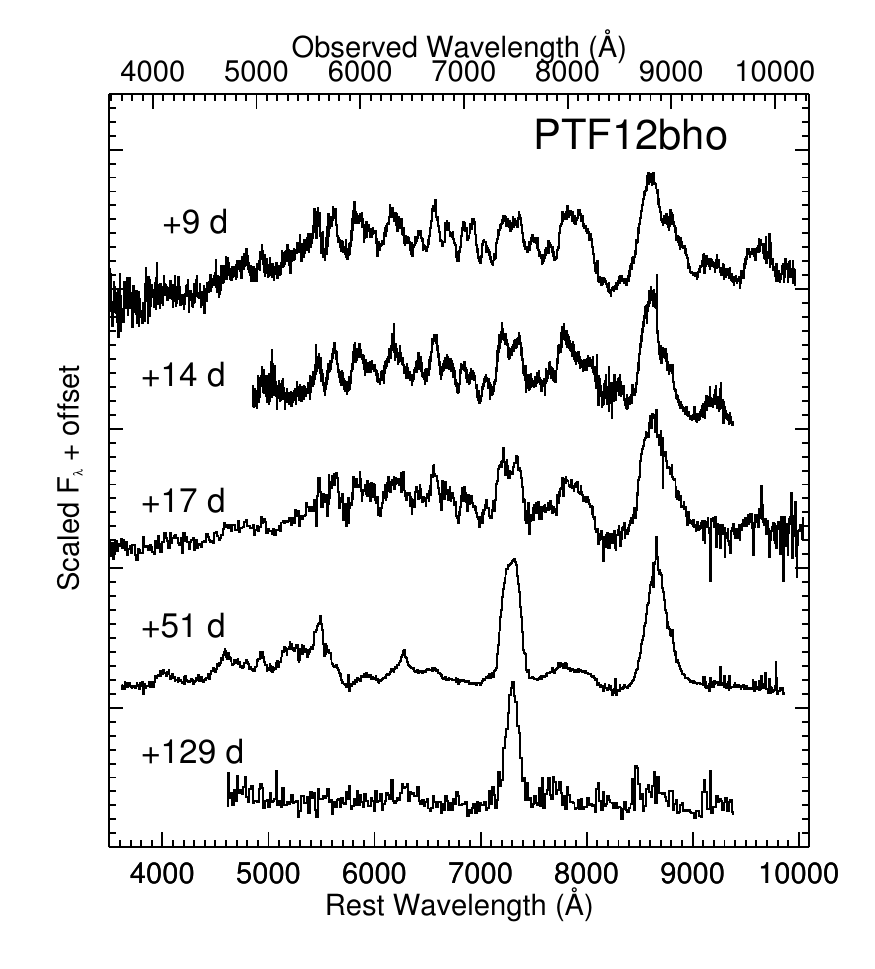}
\end{tabular}
\caption{Observed spectroscopic sequence of PTF11kmb (left) and PTF12bho (right). The phase of each spectrum with respect to the $r$-band maximum is marked. Note the early transition to the nebular phase, and the high ratio of [\ion{Ca}{2}] $\lambda\lambda$7291, 7324 compared to [\ion{O}{1}] $\lambda\lambda$6300, 6364.}
\label{fig:both_spec}
\end{figure*}

We obtained a sequence of spectra for both PTF11kmb and PTF12bho using LRIS on Keck I, the DEep Imaging Multi-Object Spectrograph (DEIMOS; \citealt{fpk+03}) on the 10-m Keck II telescope, and the Double Spectrograph (DPSP; \citealt{og82}) on the 200-in Hale telescope at Palomar Observatory (P200). The times of the spectra are marked as dashed lines in Figures~\ref{fig:11kmb_lc} and \ref{fig:12bho_lc}. Details of the spectroscopic observations are given in Table~\ref{tab:spec}, and the spectroscopic sequences for PTF11kmb and PTF12bho are shown in Figure~\ref{fig:both_spec}. The spectroscopic properties are analyzed and discussed in Section~\ref{sec:spec}.

\subsection{Host-Galaxy Photometry}
\label{sec:hst}
We obtained deep imaging of the fields of PTF11kmb using WFC3/UVIS on the \textit{Hubble Space Telescope (HST)} through program GO-13864 (PI Kasliwal). This program also covered the field of SN\,2005E \citep{pgm+10}, and we present and analyze the data on both transients here. Out of the confirmed Ca-rich gap transients so far, PTF11kmb is the farthest from any apparent host, while SN\,2005E is at the lowest redshift, making these two targets particularly interesting for searching for underlying faint sources. The details of the observations are listed in Table~\ref{tab:hst}. The individual images were corrected for charge-transfer efficiency (CTE) losses using the tools available on the WFC3 website\footnote{\url{http://www.stsci.edu/hst/wfc3/tools/cte\_tools}}, and combined using the standard Astrodrizzle package provided by STScI. Each image was drizzled to a final {\tt pixscale} of 0.02\arcsec\ pixel$^{-1}$ using a {\tt pixfrac} of 0.8.

While we do not have \textit{HST} data covering PTF12bho, there is a wealth of other survey data available thanks to its location within the Coma Cluster. In particular, the position of PTF12bho is covered by the ultra-deep $R$-band Suprime-Cam/Subaru data presented by \citet{kyy+15} and \citet{ykk+16}.

\subsection{Astrometry}
\label{sec:astrometry}
To determine the location of the transients relative to the deep \textit{HST} and Subaru images, we first run {\tt SourceExtractor}\footnote{\url{http://sextractor.sourceforge.net/}} on both the deep host images and an image containing the transient, and construct catalogs of point sources in common between the two images. We then use IRAF's {\tt ccmap} task to compute the astrometric tie between the two images and shift the SN image to the \textit{HST} reference frame, achieving a combined astrometric uncertainty of 84~mas in the relative position of PTF11kmb and 76~mas in the position of PTF12bho. For SN\,2005E, there are not enough sources in common between the deep and small field of view \textit{HST} image and the larger, shallower images containing the SN, so we use an image from the Sloan Digital Sky Survey (SDSS) as an intermediate step. The (combined) uncertainty in the astrometric tie for SN\,2005E is 98~mas.

\subsection{Host-Galaxy Spectroscopy}
PTF11kmb is the Ca-rich transient discovered at the largest projected distance to the presumed host galaxy (NGC\,7265) to date: 150~kpc. NGC\,7265 is a member of a galaxy group, however, and several extended sources near PTF11kmb lack redshifts in NED. Thus, we carried out two spectroscopic mask observations of the field around PTF11kmb using DEIMOS on the Keck-II telescope, in order to assess whether any of the sources closer to PTF11kmb were at a redshift consistent with the group and thus potential host galaxies at smaller offsets. The redshifts and coordinates of objects confirmed to be galaxies are listed in Table~\ref{tab:11kmbgal}, and are discussed in Section~\ref{sec:offsets}.

\section{Two New Ca-rich Gap Transients: \\ PTF11kmb and PTF12bho}
\label{sec:gap}

In this section, we analyze the properties of PTF11kmb and PTF12bho, and show that they are members of the class of Ca-rich gap transients.

\subsection{Light-Curve Properties}

The individual light curves of PTF11kmb and PTF12bho are shown in Figures~\ref{fig:11kmb_lc} and \ref{fig:12bho_lc}, respectively. The peak luminosities of PTF11kmb and PTF12bho are $r = -15.5$~mag and $r = -16.0~{\rm mag}$ (respectively), comparable to previous Ca-rich gap transients. The $r$-band light curves are the best sampled, and we fit a low-order polynomial to the light curves near peak brightness to determine a best-fit time of peak: MJD 55784.9 for PTF11kmb and MJD 55992.4 for PTF12bho. All phases reported are relative to these peak times.

The light-curve rise is best sampled for PTF12bho, where both pre-explosion limits as well as a $t^2$ fit to the rising portion of the light curve in flux space yield a rest-frame rise time to $r$-band maximum of 12~days. The rise is not quite as well constrained for PTF11kmb owing to a gap in coverage before the first detection: the last upper limit is 18~days and the first detection 11~days prior to peak; fitting a $t^2$ curve gives a rest-frame rise time of 15~days. We can use these measured rise times, as well as the velocity at peak determined from spectroscopy (Section~\ref{sec:photspec}), to make a simple estimate of the ejecta masses from $M_{\rm ej} \propto v t^2$ \citep{arn82}. Scaling from a normal SN~Ia ($M_{\rm ej} = 1.4~{\rm M}_{\odot}$, $v = 11,000~{\rm km~s}^-1$, $t_{\rm rise} = 17.4$~days), we get approximate ejecta masses $M_{\rm ej} \approx 0.6 - 1.0~{\rm M}_{\odot}$ for PTF11kmb and $M_{\rm ej} \approx 0.4~{\rm M}_{\odot}$ for PTF12bho. 

\begin{figure}
    \centering
    \includegraphics[width=3.5in]{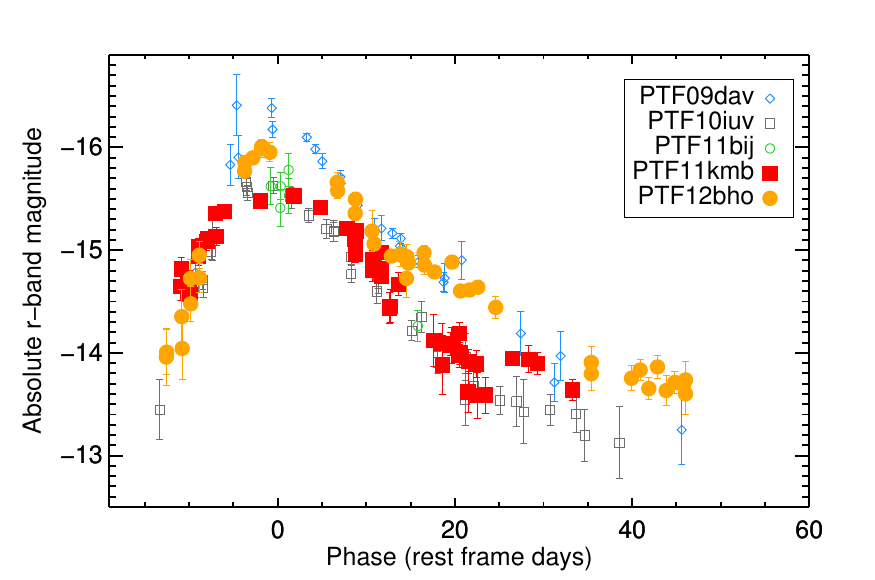}
    \caption{The $r$-band light curves of PTF11kmb and PTF12bho compared to the Ca-rich gap transients from PTF presented by K12. PTF11kmb and PTF12bho have very similar light curves to the previous objects, with relatively low luminosities and short rise times compared to typical SNe.}
    \label{fig:lccomp}
\end{figure}

Figure~\ref{fig:lccomp} shows the $r$-band light curves of PTF11kmb and PTF12bho compared to the three previous PTF Ca-rich gap transients from K12. PTF11kmb and PTF12bho have very similar light curves to the previous events, in terms of both their peak luminosities and their rise and decline timescales. Thus, we conclude that PTF11kmb and PTF12bho satisfy the light-curve criteria of K12 for Ca-rich gap transients: low peak luminosities and rapid photometric evolution.

\subsection{Spectroscopic Properties}
\label{sec:spec}

\subsubsection{Photospheric-Phase Spectra}
\label{sec:photspec}
\begin{figure}
\centering
\includegraphics[width=3.4in]{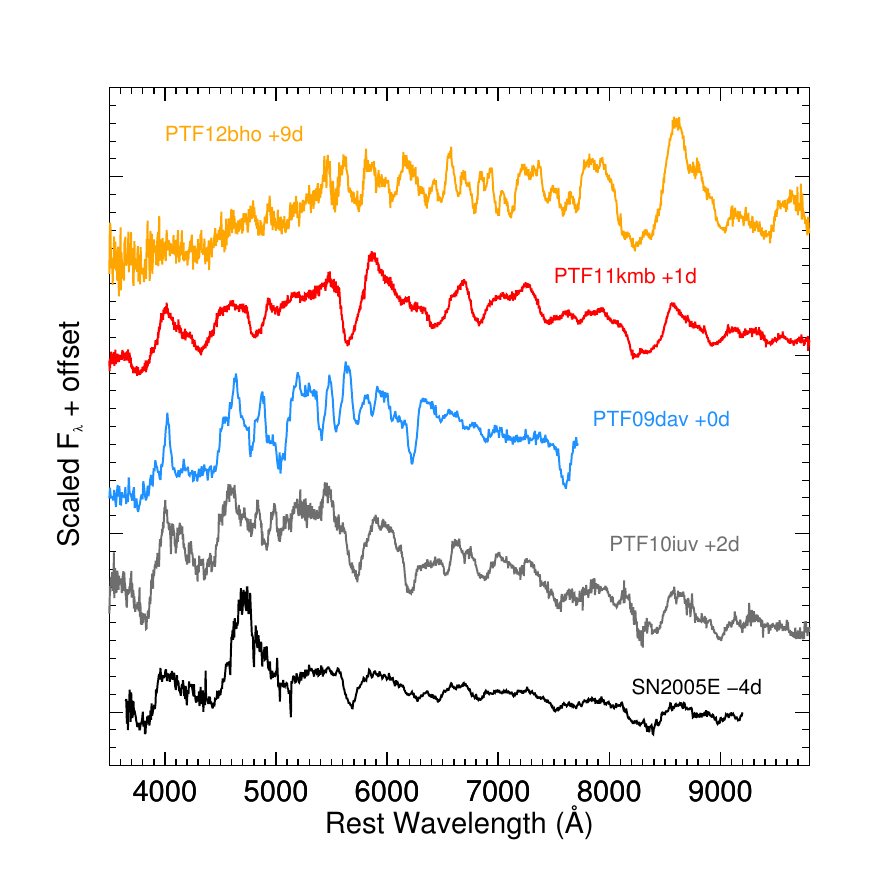}
\caption{Photospheric-phase spectra of PTF11kmb and PTF12bho, compared to other Ca-rich gap transients. The photospheric-phase spectra of this class are diverse, though generally they show velocities comparable to those of other SNe. Many exhibit He features --- the photospheric spectrum of PTF11kmb is an excellent match to that of a Type Ib SN. The spectrum of PTF12bho is very complicated, and we analyze it in more detail in Figure~\ref{fig:synow}.}
\label{fig:phot_spec}
\end{figure}

We show example photospheric-phase spectra of PTF11kmb and PTF12bho in Figure~\ref{fig:phot_spec}, compared to the Ca-rich gap transients from K12. PTF11kmb was originally classified as a Type Ib SN \citep{gxb+11} based on the  strong He features in this spectrum. Helium features in the photospheric spectrum are a common (if not defining) property of Ca-rich gap transients, and were also seen in SN\,2005E, PTF10iuv, and SN\,2007ke. Using the minimum of the He lines, we measure a photospheric velocity of $\sim 11,000~{\rm km~s}^{-1}$ for PTF11kmb.

The photospheric spectrum of PTF12bho is more complicated and warrants further investigation. We use the spectral synthesis code {\tt SYN++} \citep{synow} to identify as many of the lines as possible, as well as to get an estimate of the photospheric expansion velocity. {\tt SYN++} calculates a synthetic spectrum by considering an optically thick pseudo-photosphere surrounded by an extended line-forming region, using a number of simplifying assumptions including spherical symmetry, local thermodynamic equilibrium (LTE), and only thermal excitations. The fit is constrained by matching both the absorption profiles and the relative strength of different features.

\begin{figure}
    \centering
    \includegraphics[width=3.5in]{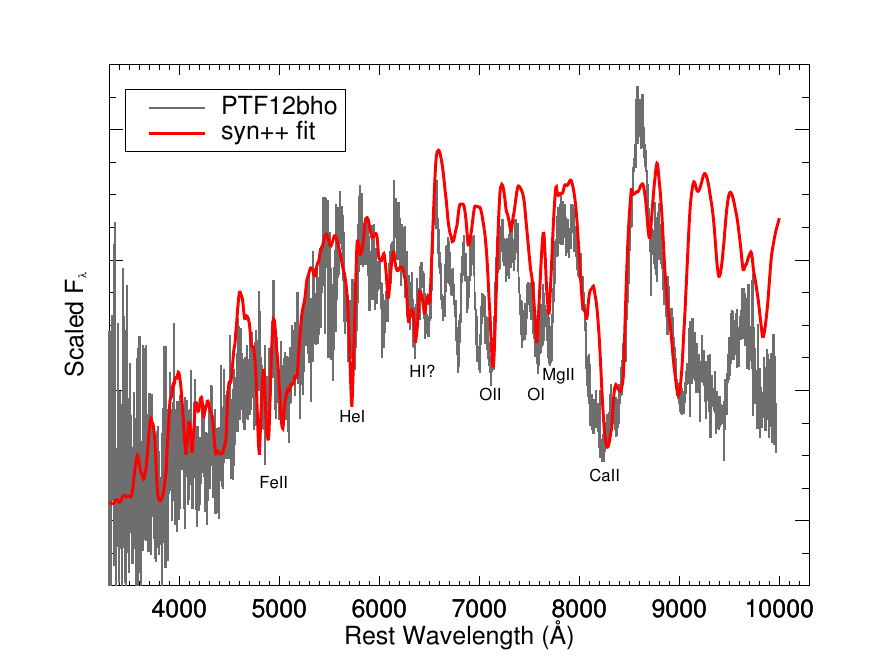}
    \caption{{\tt SYN++} fit to the +9~day spectrum of PTF12bho. The spectrum is very complex and a number of features are still unidentified. The main species we can securely identify are \ion{Ca}{2}, \ion{O}{1}, \ion{Fe}{2}, \ion{Mg}{2}, and \ion{He}{1}.}
    \label{fig:synow}
\end{figure}

Figure~\ref{fig:synow} shows the resulting {\tt SYN++} fit for the $+9~{\rm day}$ PTF12bho spectrum, with the main features marked. The fit shown has a blackbody temperature of 4500~K and velocities in the range of 6000--10,000~km~s$^{-1}$. We consider \ion{Ca}{2}, \ion{O}{1}, \ion{Mg}{2}, \ion{Fe}{2}, and \ion{He}{1} as reasonably secure identifications based on multiple lines of the same species seen in the spectrum. The relative line strengths of \ion{He}{1} are not perfectly reproduced, but this is expected as it would require non-LTE calculations which are not included in {\tt SYN++} (e.g., \citealt{luc91,bhn+96,bbk+02}).

A number of other species were attempted and can neither be ruled out nor securely identified, as they generally produce only one strong feature in the spectrum. These possible species include \ion{C}{2}, \ion{O}{2}, and \ion{Ca}{1}. Both \ion{Co}{2} and \ion{Ti}{2} are possible, and mainly contribute in bringing the flux down in the blue. \ion{Al}{2} can match the feature at 6780~\AA\,, but only if moving at 12,000~km~s$^{-1}$, significantly above the photospheric velocity. \ion{Sc}{2} (which was seen in PTF09dav; \citealt{skn+11}) was considered for the feature at 5540~\AA\,, but the corresponding feature at 6050~\AA\, is not well matched, and the expected strong line at 5400~\AA\, is not seen in the spectrum. Interestingly, we also cannot rule out \ion{H}{1}, which matches the feature at 6350~\AA\, if moving at a velocity of $\sim 10,000$~km~s$^{-1}$, comparable to other species in the fit. 

A number of features are still unidentified (5540~\AA\,, 6640~\AA\,, 6780~\AA\,, 6700~\AA\,, 7430~\AA\,, 8080-8220~\AA\,, 8770~\AA\,, 9400~\AA\,). The failure of {\tt SYN++} to find matches for these features is likely due to a combination of factors: problems with the (LTE) line strengths assumed in the model, forbidden lines, or lines moving at significantly higher velocities than the photosphere. The uncertainty in the precise redshift of PTF12bho owing to the lack of an unambiguously identified host galaxy also adds to the difficulty. While a more thorough exploration of the spectral sequence of PTF12bho would be very interesting for shedding light on this object, we consider it outside the scope of this paper.

We note that when K12 were attempting to empirically define the class of Ca-rich gap transients, they allowed for diversity in the photospheric-phase spectra, and did not (for example) require the detection of He. Without this allowance, an object like PTF09dav which was originally reported as a subluminous SN~Ia would not have been included \citep{skn+11}. PTF12bho adds to this diversity, with a photospheric spectrum that is unique among the Ca-rich gap transients discovered to date. Understanding this diversity will be important in determining the physical phenomenon giving rise to the Ca-rich gap transients, including whether objects like PTF09dav and PTF12bho are truly part of the same class of objects.

\subsubsection{Nebular-Phase Spectra}

\begin{figure}
\centering
\includegraphics[width=3.4in]{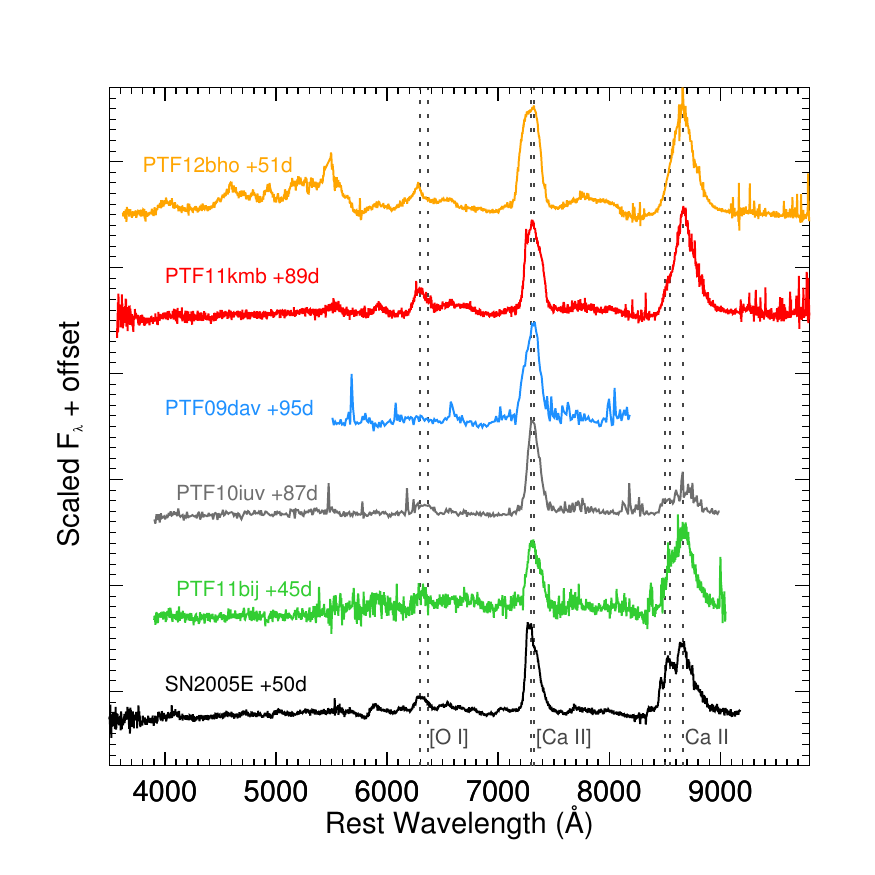}
\caption{Nebular-phase spectra of PTF11kmb and PTF12bho, compared to those of other Ca-rich gap transients. PTF12bho shows a mix of photospheric and nebular features in its spectrum at 52~days past peak; the same was true for SN\,2005E. Note the high ratio of Ca to O, particularly the strong [\ion{Ca}{2}] $\lambda\lambda$7291, 7324 emission line which is characteristic of this class of transient. }
\label{fig:neb_spec}
\end{figure}

As with other Ca-rich gap transients, PTF11kmb and PTF12bho transition to the nebular phase on relatively short timescales. A $+51$~day spectrum of PTF12bho and a $+89$~day spectrum of PTF11kmb are shown in Figure~\ref{fig:neb_spec}, again compared to nebular spectra of other members of this class. While the PTF12bho spectrum still exhibits some photospheric features (which was also the case for SN\,2005E at a comparable phase), both spectra are dominated by nebular \ion{Ca}{2} emission. Note in particular the strong forbidden [\ion{Ca}{2}] $\lambda\lambda$7291, 7324 emission line, and the comparatively weak [\ion{O}{1}] $\lambda\lambda$6300, 6363 emission. This is one of the defining characteristics of the Ca-rich gap transient class --- while other SN types may also show strong [\ion{Ca}{2}] emission, the ratio of [\ion{Ca}{2}] to [\ion{O}{1}] is typically much lower (see e.g. Figure~4 of K12).

We also note that at the phases displayed here, PTF11kmb and PTF12bho also exhibit strong emission in the permitted \ion{Ca}{2} triplet --- our series of spectra (Figure~\ref{fig:both_spec}) show that the strength of this line decreases with time, while the forbidden [\ion{Ca}{2}] line stays strong compared to [\ion{O}{1}] in all of our nebular spectra. Thus, using the strength of [\ion{Ca}{2}] relative to [\ion{O}{1}] as one of the defining characteristics of Ca-rich gap transients does not seem to depend strongly on the exact phase of the spectrum.

Combining the photometric and spectroscopic information, then, we have shown that PTF11kmb and PTF12bho both satisfy the five characteristics of Ca-rich gap transients: (1) low peak luminosities, (2) rapid photometric evolution, (3) normal photospheric-phase velocities, (4) early evolution to the nebular phase, and (5) nebular spectra with a high ratio of Ca compared to O. Thus, we conclude that both objects are unambiguous members of the class of Ca-rich gap transients as described by K12.

\section{Galaxy Environments of Ca-Rich Gap Transients}
\label{sec:gal}
 A peculiar feature of Ca-rich gap transients is that they tend to be found at large offsets from any apparent host (K12). This is true also of PTF11kmb and PTF12bho: no underlying host is visible in the PTF reference images for either transient. PTF11kmb stands out in particular from being at least 80~kpc from the closest host candidate with a redshift in NED. The closest galaxy to PTF12bho is SDSS~J130109.43+280159.2, at a projected distance of 48~kpc. In this section, we analyze the host environments of PTF11kmb and PTF12bho in detail, and discuss them in the context of the sample of Ca-rich gap transients.

\subsection{Limits on In-Situ Formation}
\label{sec:limits}

\begin{figure*}
\centering
\begin{tabular}{ccc}
\includegraphics[width=5.7cm]{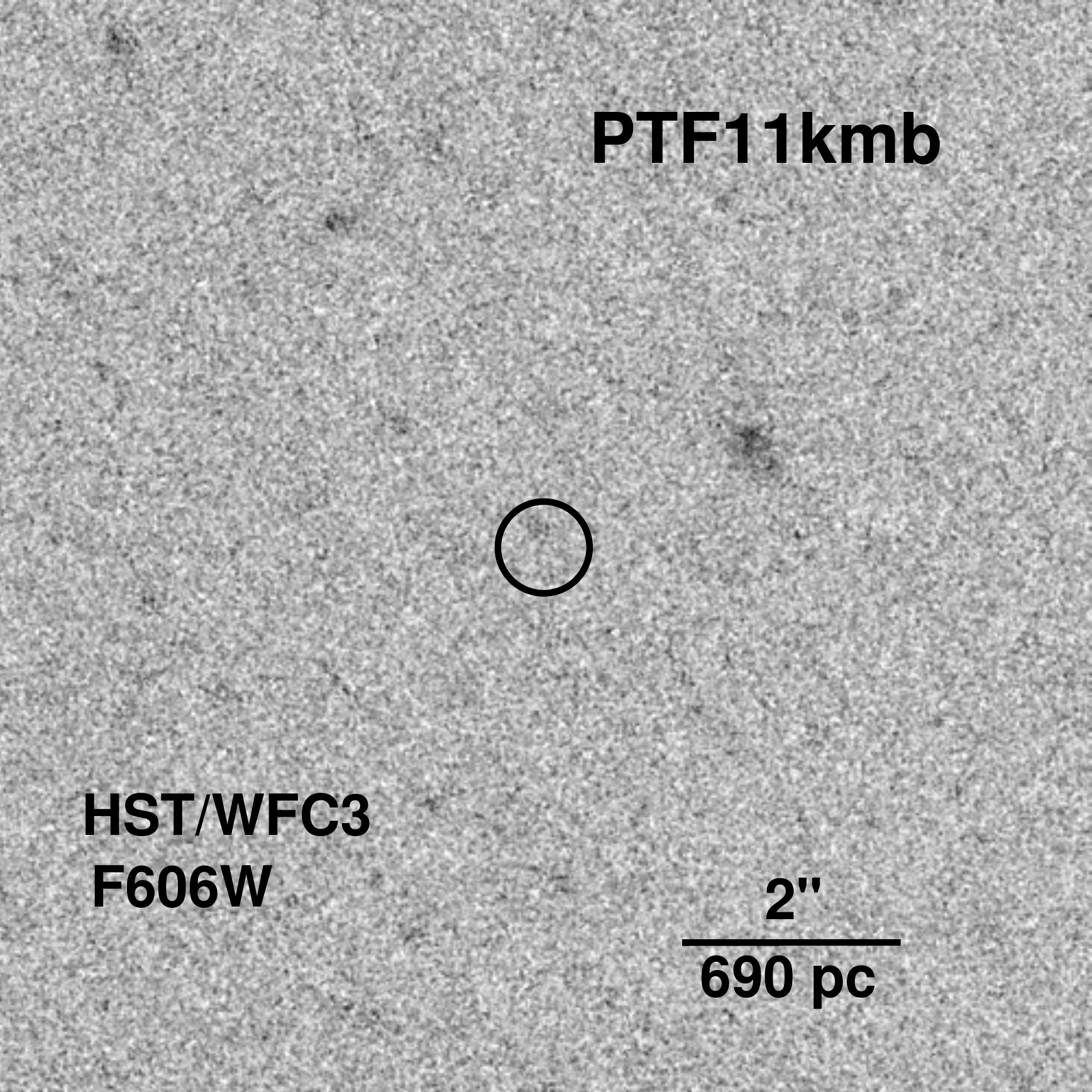} & \includegraphics[width=5.7cm]{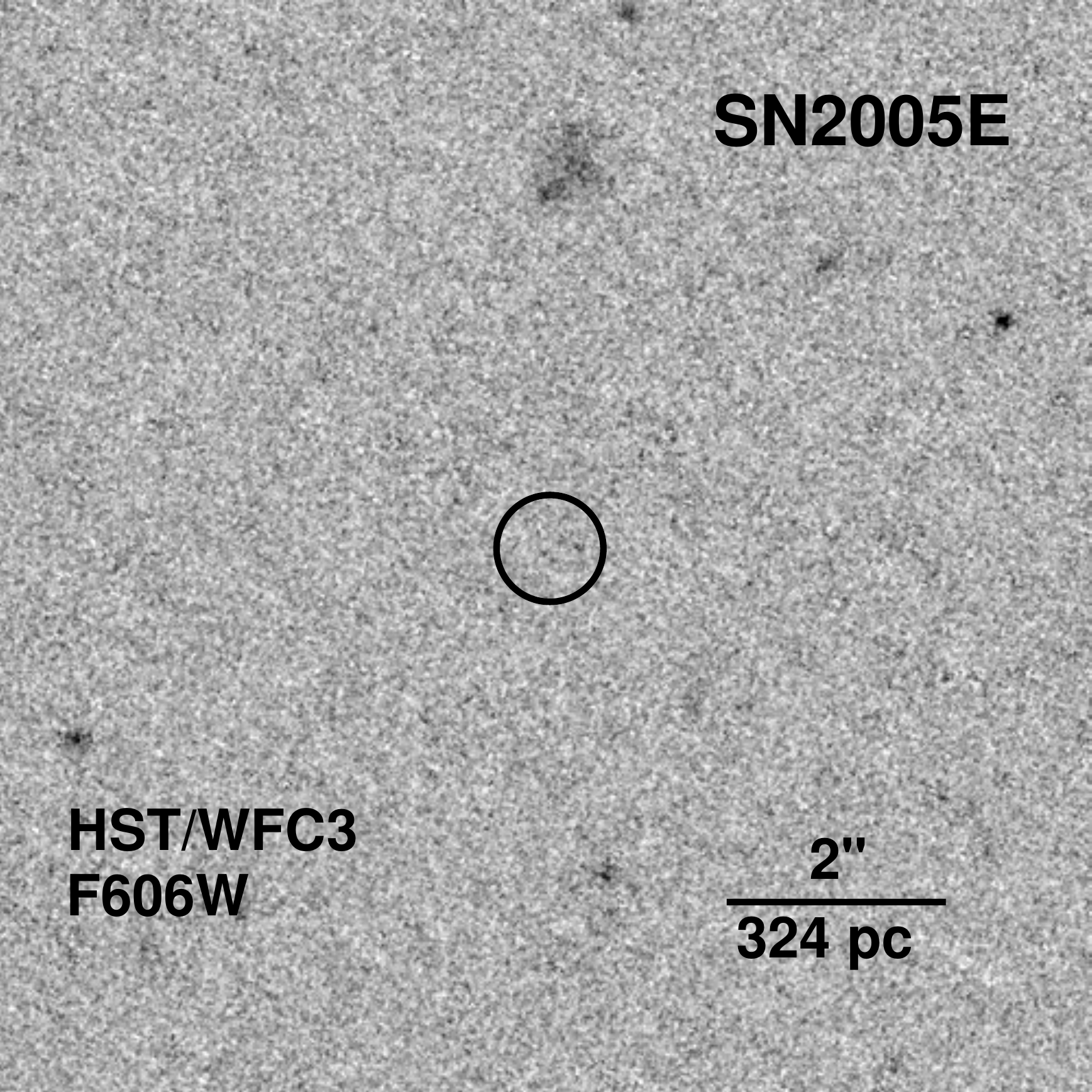} & \includegraphics[width=5.7cm]{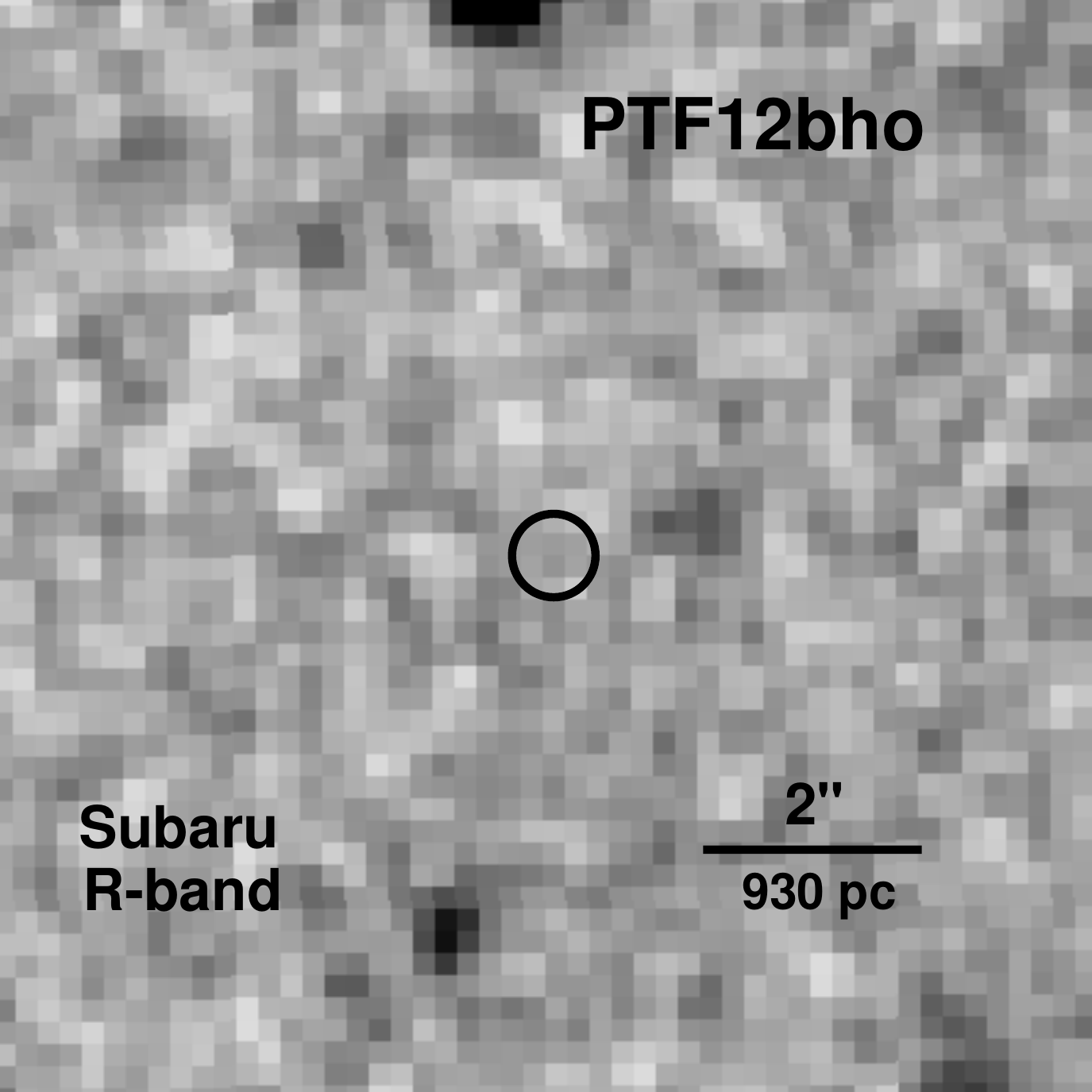}
\end{tabular}
\caption{{\it HST} imaging of the fields around PTF11kmb (left) and SN\,2005E (middle), and deep Subaru imaging of the field around PTF12bho (right). All images are 10\arcsec\ $\times$ 10\arcsec, and oriented with north up and east to the left. The locations of the transients are noted by the circles, corresponding to the $5\sigma$ uncertainty radius. No sources are detected at the positions of SN\,2005E and PTF12bho, while there is a marginally detected source within the uncertainty radius of PTF11kmb.}
\label{fig:deep_img}
\end{figure*}

As described in Section~\ref{sec:hst}, we have deep imaging of the locations of PTF11kmb and SN\,2005E from \textit{HST}, and of PTF12bho from Subaru. The images at the transient positions are shown in Figure~\ref{fig:deep_img}. Each cutout is 10\arcsec $\times$ 10\arcsec, and the circles correspond to the $5\sigma$ uncertainty radius in the transient position from the astrometric tie (Section~\ref{sec:astrometry}). No underlying sources are seen at the positions of SN\,2005E or PTF12bho, to upper limits $m_{606W} > 28.9~{\rm mag}$ and $m_R >27.0~{\rm mag}$, respectively. The upper limit for SN\,2005E corresponds to an absolute magnitude $M_{606W} > -4.0~{\rm mag}$, thus ruling out the majority of the globular cluster luminosity function \citep{jmc+07}; for PTF12bho the upper limit corresponds to an absolute magnitude $M_r > -8.0~{\rm mag}$. Similarly deep limits also exist for the Ca-rich gap transients SN\,2012hn ($M_R > -5.6~{\rm mag}$; \citealt{llc+14}) and SN\,2007ke ($M_{606W} > -6.6~{\rm mag}$; \citealt{llj+16}), indicating that the majority of Ca-rich gap transients likely did not form at their explosion sites. 

The case of PTF11kmb is slightly more complicated: here, there is a marginally significant source within the $5\sigma$ position error circle. Aperture photometry yields an apparent magnitude $m = 28.3 \pm 0.15~{\rm mag}$. If at the same redshift as PTF11kmb, this corresponds to an absolute magnitude of $-6.0~{\rm mag}$, which would be consistent with a globular cluster, though on the faint end of the luminosity function. The putative source is too faint for spectroscopy to assess whether it is at the same redshift as PTF11kmb, and other similarly faint features in the image appear streak-like and likely artifacts produced by CTE correction problems. Deeper imaging, as well as imaging in multiple filters, can establish whether the marginal source is real, and whether its colors are best explained by a globular cluster or a background/foreground source.

We also caution that at these faint magnitudes, the sky density of background sources is high. Following \citet{ber10} and \citet{bkd02}, based on the source density distribution observed in surveys like the Hubble Ultra Deep Field \citep{bsk+06}, the probability of a 28.3~mag source within our $5\sigma$ uncertainty radius being a random background galaxy is about $7\%$. While the source appears to be extended, the signal-to-noise ratio is low, so we cannot completely rule out a foreground star, though we consider it less likely: for a source this faint to be within the Galaxy, it would have to be either a late-M dwarf or a white dwarf (absolute magnitude 11--12) far out in the halo. Given the marginal nature of the source and the relatively high likelihood of chance coincidence, we conclude that a globular cluster origin for PTF11kmb cannot be ruled out.

\subsection{Offset Distribution}
\label{sec:offsets}
\begin{figure*}
    \centering
    \includegraphics[width=7in]{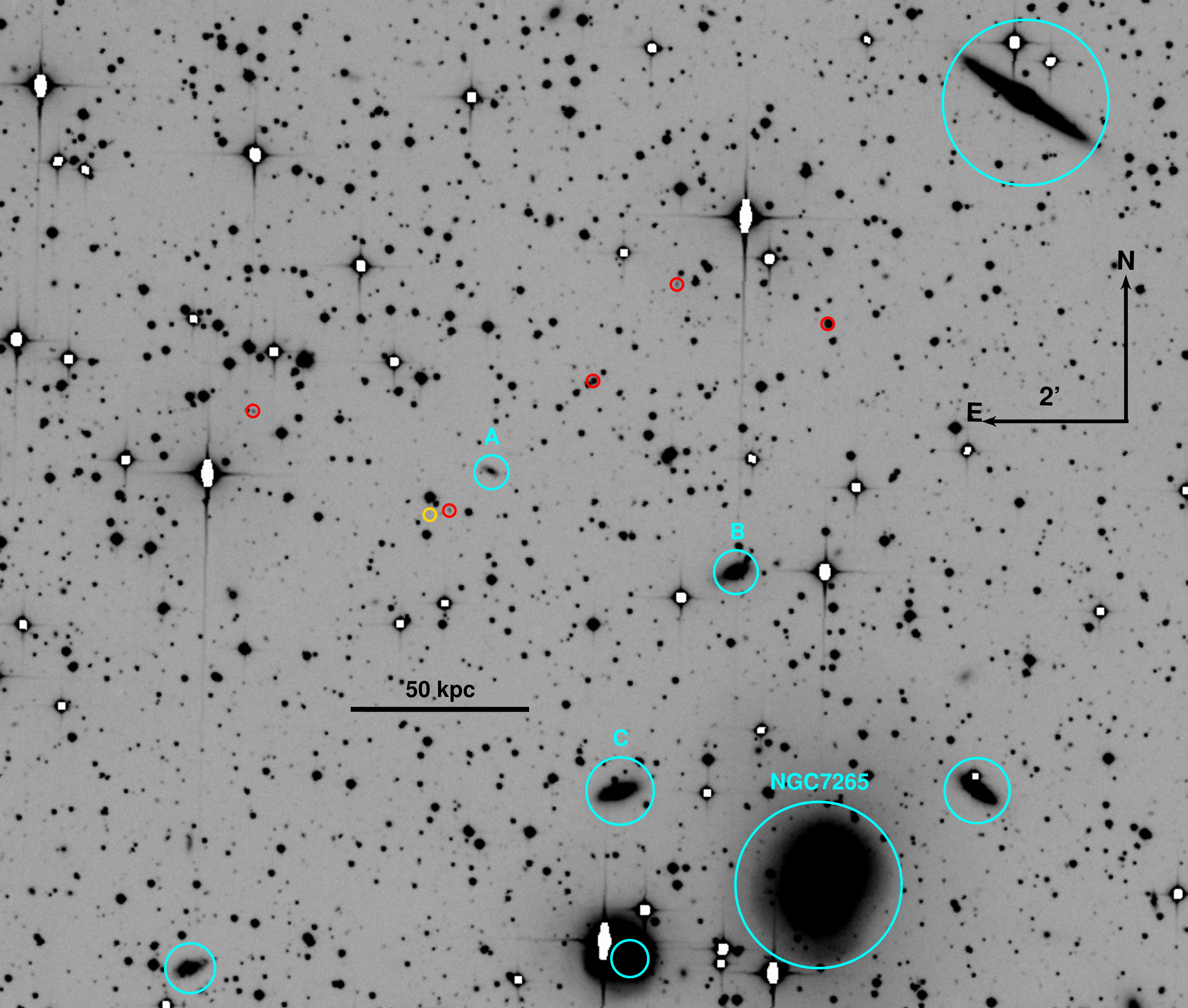}
    \caption{PTF $R$-band reference image of the field around PTF11kmb. The yellow circle marks the position of the transient. Objects circled in cyan are galaxies with redshifts consistent with the galaxy group. The red circles mark objects that were found to be background galaxies from our spectroscopic mask. All other objects identified from the mask turned out to be foreground stars. }
    \label{fig:11kmb_img}
\end{figure*}

Identifying the true host galaxies of either PTF11kmb or PTF12bho is ambiguous. \citet{fol15} argued that NGC\,7265 is the most likely host of PTF11kmb: while offset by $\sim 150$~kpc in physical units, it is the closest galaxy in terms of isophotal radii and the closest in redshift to PTF11kmb out of the galaxies with a redshift in NED. However, our spectroscopy shows that there are at least two more galaxies that are likely associated with the galaxy group and with smaller projected offsets: GALEXASC~J222249.41+361811.7 at 1\arcmin\ northwest and $z=0.021$, labeled ``A'' in Figure~\ref{fig:11kmb_img}, and 2MASX~J22223308+3616504 at 4.2\arcmin\ west-southwest at $z=0.0179$, labeled ``B.'' We note that unlike the other nearby group galaxies, GALEXASC~J222249.41+361811.7 exhibits strong H$\alpha$ emission, indicating ongoing star formation. Importantly, no galaxy at a consistent redshift with PTF11kmb was found closer than 1\arcmin\ -- while a source 16.2\arcsec\ west of PTF11kmb was revealed to be a galaxy (circled in red), we find that it is background at $z=0.166$, and therefore unrelated to the transient.

GALEXASC~J222249.41+361811.7 (``A'') is the closest potential host galaxy to PTF11kmb in terms of both projected distance and host-normalized projected distance ($21~{\rm kpc}$ and $\sim 16$ galaxy half-light radii, respectively), though offset by $\sim 1200~{\rm km~s}^{-1}$ from the redshift of $z = 0.017$ derived from the SN spectrum. NGC\,7265 at $150~{\rm kpc}$ is the second closest in terms of host-normalized distance ($\sim 20$~half-light radii), and at a redshift consistent with that derived from the SN ($z = 0.0169$). The galaxies labeled ``B'' and ``C'' in Figure~\ref{fig:11kmb_img} are closer than NGC\,7265 in terms of projected distance, though not in host-normalized distance. For the purposes of measuring offsets, we assume NGC\,7265 is the host of PTF11kmb as it is closer in redshift and offset a similar host-normalized distance compared to GALEXASC~J222249.41+361811.7. We note that our results are not significantly influenced by this choice, however: either way, PTF11kmb is significantly offset both in physical and host-normalized units, and the statistics comparing Ca-rich gap transients to other populations are not affected by instead adopting GALEXASC~J222249.41+361811.7 as the host.

\begin{figure*}
    \centering
    \begin{tabular}{cc}
    \includegraphics[width=3.5in]{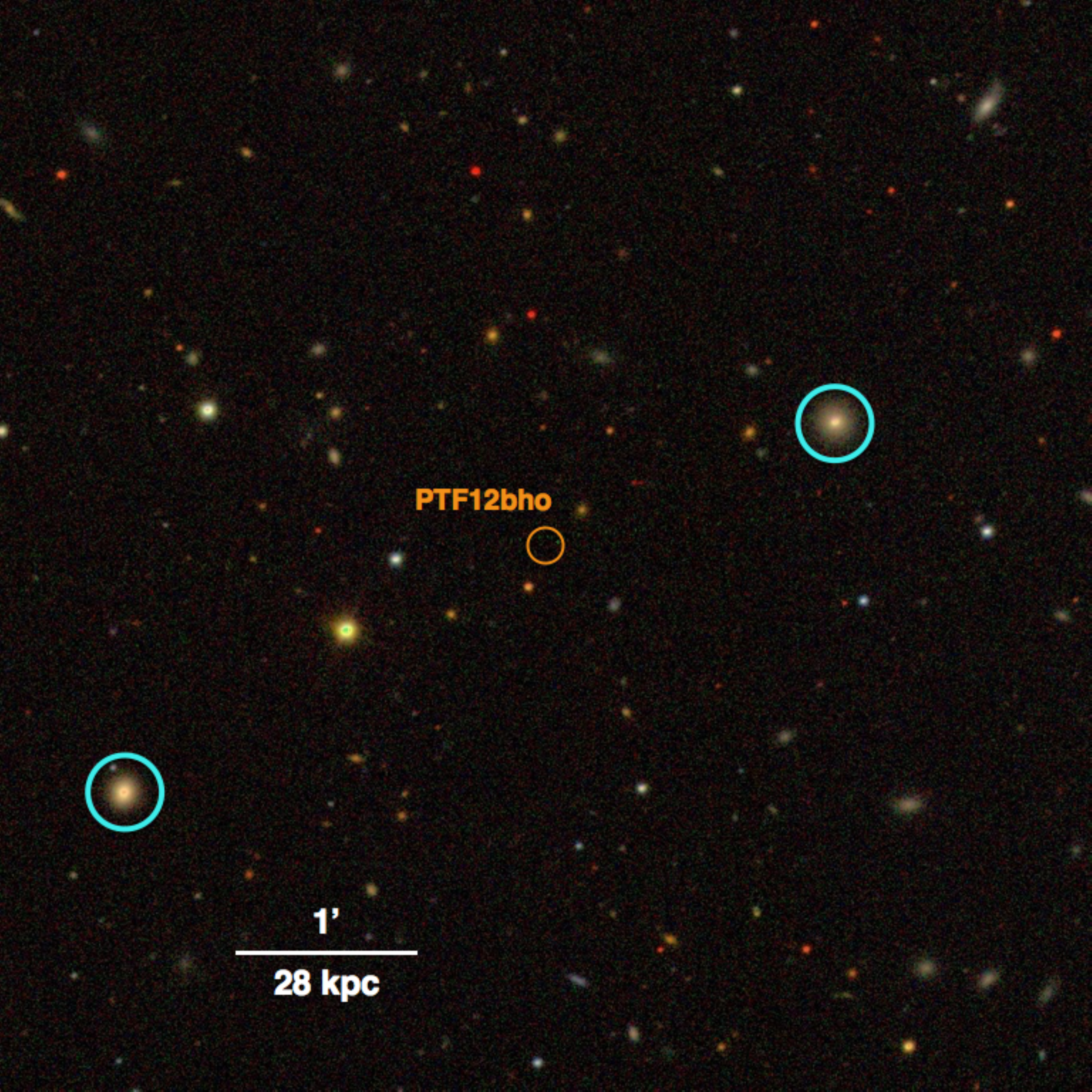} &
    \includegraphics[width=3.5in]{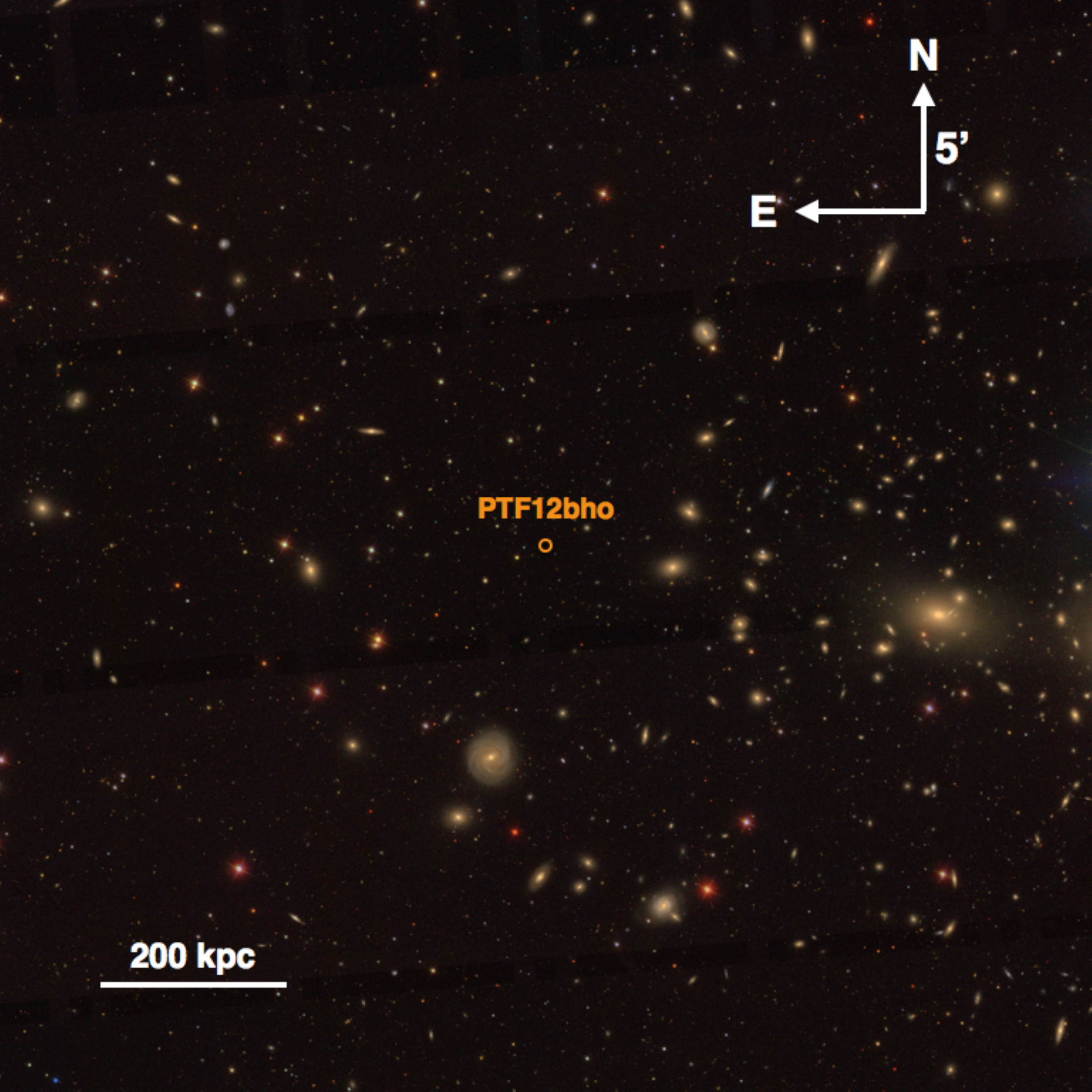}
    \end{tabular}
    \caption{Left: $6\arcmin \times 6\arcmin$ SDSS $gri$ composite image of the field around PTF12bho. The two closest galaxies having redshifts consistent with the Coma Cluster are circled. Right: $42\arcmin \times 42\arcmin$ wide-field view showing the position of PTF12bho within the Coma Cluster.}
    \label{fig:12bho_img}
\end{figure*}

Similarly, PTF12bho is 1.73\arcmin\ (48~kpc) away from SDSS~J130109.43+280159.1, the nearest galaxy with a measured redshift, and which we nominally adopt as the host. On the other hand, being located in the intra-cluster medium of the Coma Cluster, there are a number of potential host galaxies at a compatible redshift. Figure~\ref{fig:12bho_img} shows the field around PTF12bho with the two closest galaxies with compatible redshifts circled (left), as well as a zoom-out of the field showing PTF12bho's location within the Coma Cluster (right). 

\begin{figure*}
    \centering
    \begin{tabular}{cc}
    \includegraphics[width=3.4in]{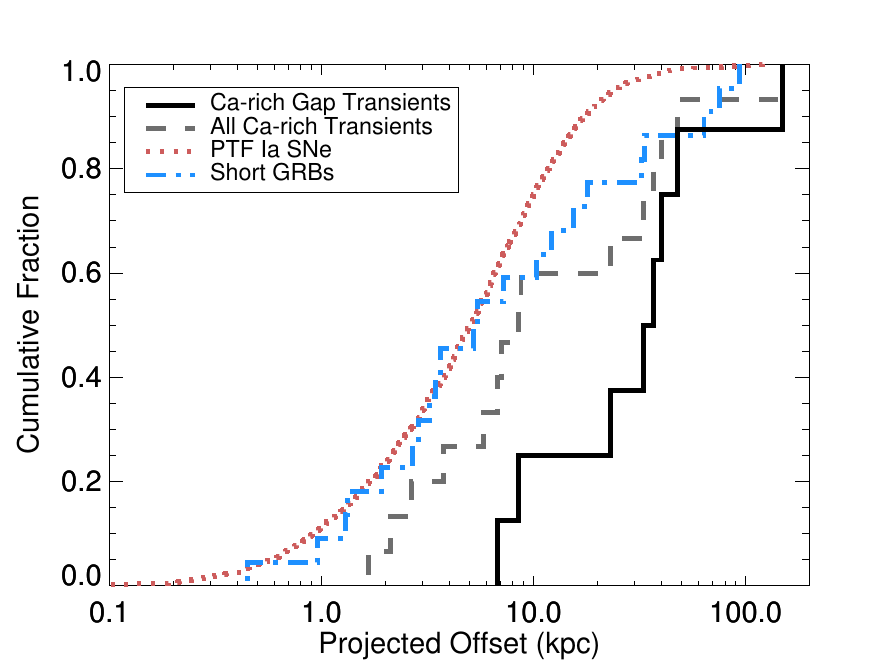} & \includegraphics[width=3.4in]{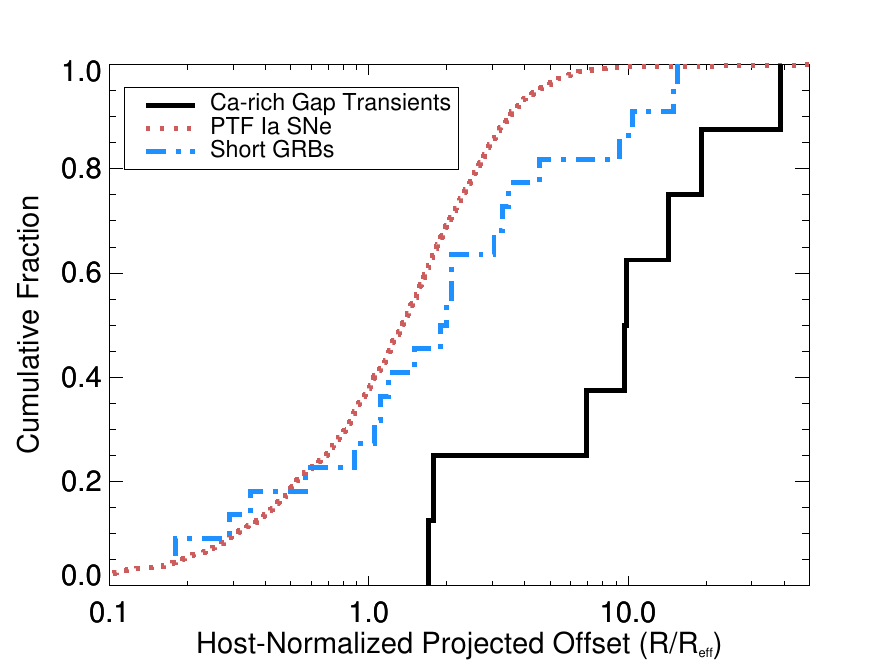}
    \end{tabular}
    \caption{Offset distribution of Ca-rich gap transients, shown both in physical units (left) and normalized by the host galaxies' half-light radius (right). The solid black line shows the Ca-rich gap transients; for comparison we also show the offset distributions of short GRBs from the sample of \citet{fbf10,fb13} (blue dot-dashed line) and SNe~Ia from the PTF survey (red dotted line). The offset distribution of the Ca-rich gap transients is significantly more extreme than either of these transient classes. The gray dashed line in the left plot shows the offset distribution for all ``Ca-rich transients'', i.e. also including the candidates that lack photometric information. This distribution is less extreme, and we discuss potential reasons for the discrepancy, including selection effects in PTF, in Section~\ref{sec:selection}.}
    \label{fig:offsets}
\end{figure*}

With these host associations, Figure~\ref{fig:offsets} shows the resulting offset distribution for the known Ca-rich gap transients, compared to two other transient populations believed to have old, binary progenitors: Type Ia SNe (showing the PTF sample; L.~Hangard et al., in prep.), and short-duration gamma-ray bursts (sGRBs; \citealt{fbf10,fb13}).  The left panel displays the physical projected offsets, while the right shows offsets normalized by the hosts' half-light radii. In either view, the offset distribution of Ca-rich gap transients is extreme, with all of the objects in the confirmed sample found at offsets greater than one half-light radius. The differences with both SNe~Ia and sGRBs are statistically significant: $p = 0.0004$ and $p = 0.04$ (respectively), as determined by a Kolmogorov-Smirnov test.

\subsubsection{Survey Selection Effects}
\label{sec:selection}
It is potentially worrying that all of the PTF Ca-rich gap transients to date have been found at offsets $> 20~{\rm kpc}$, while all the (candidate) Ca-rich gap transients from other surveys were found at offsets $< 10~{\rm kpc}$ with the exception of SN\,2005E at 23~kpc. This is also illustrated by the gray line in Figure~\ref{fig:offsets}, which shows the distribution of offsets including the candidate Ca-rich transients that lack light-curve information. The best way to quantify the extent to which PTF's detection efficiency for fast and faint transients is affected by the underlying galaxy surface brightness would be to inject fake transients and run a recovery analysis. Such an analysis is outside the scope of this paper, but will be presented by C.~Frohmaier et al. (2017, in prep.). 

As a simpler test of whether PTF is missing a large fraction of faint transients on top of galaxies, we search the PTF database for all objects that were classified as having ``gap'' luminosities between 2009 and 2012. We find 17 such objects in total, including the five Ca-rich gap transients PTF09dav, PTF10iuv, PTF11bij, PTF11kmb, and PTF12bho. Our search revealed one additional object, PTF10hcw, as a candidate Ca-rich gap transient: it was reported as a SN~Ib based on the single spectrum taken \citep{gaa+10}, but also shows unusually strong Ca emission similar to that of SN\,2005E. While we lack both a well-sampled light curve and a nebular spectrum, and so cannot confirm that it was a Ca-rich gap transient, we show the light curve and spectrum in Appendix~\ref{sec:10hcw}. We note that PTF10hcw was found in the outskirts of NGC\,2639, offset 25.7\arcsec\ (5.9~kpc) from the nucleus, and so is an example of a possible PTF Ca-rich gap transient detected on top of its host galaxy.

The 11 remaining ``gap'' objects include PTF10bhp (a ``SN .Ia'' candidate; \citealt{kkg+10}), one faint Type Ic SN, and 9 objects with hydrogen emission that are likely either LBV outbursts or luminous red novae (N. Blagorodnova et al., in prep.). Notably, these objects exhibit both comparable luminosities and similar or faster timescales than the Ca-rich gap transients, and all 11 of these transients were detected on top of their host-galaxy disks. Thus, we show that PTF is capable of detecting fast and faint transients on bright galaxy backgrounds. While the detection efficiency is likely smaller in such regions of high surface brightness, the relative lack of Ca-rich gap transients found at small offsets in PTF is unlikely to be caused solely by a bias against being able to detect them. 

With the exception of SN\,2012hn, which was discovered by the Catalina Real-Time Transients Survey \citep{ddm+09} and followed up by PESSTO \citep{vyt+14}, the remainder of the candidate Ca-rich gap transients were found by the Lick Observatory Supernova Survey (LOSS; \citealt{flt+01}). LOSS is a galaxy-targeted survey, so an alternative explanation for the discrepancy could be that LOSS is missing transients at large offsets, or at least is less efficient than PTF in discovering and monitoring them (although \citet{fol15} argued that the LOSS field of view is sufficiently large to have been able to detect the objects reported by K12). A final possibility is that not all of the candidates from LOSS belong to the class of Ca-rich gap transients: \citet{vyt+14} point out that strong [\ion{Ca}{2}] nebular emission can also be seen in faint core-collapse SNe \citep{pzt+04} and in other peculiar Type I explosions such as SN\,2008ha \citep{vpc+09,fcf+09}. The offset discrepancy can likely be explained by a combination of small-number statistics, contamination by other transients with strong [\ion{Ca}{2}] in their nebular spectra, and a potential survey bias in LOSS against identifying and following up objects found far away from the targeted galaxy as SNe.

\subsection{Group and Cluster Environments}
\label{sec:vrad}
\begin{figure*}
    \centering
    \begin{tabular}{ccc}
    \includegraphics[width=6cm]{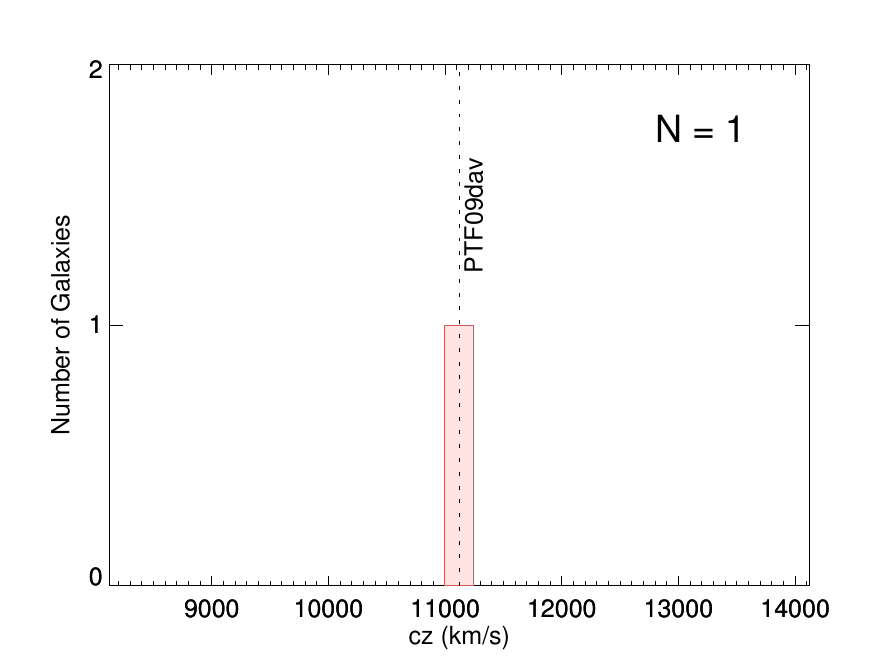} &
    \includegraphics[width=6cm]{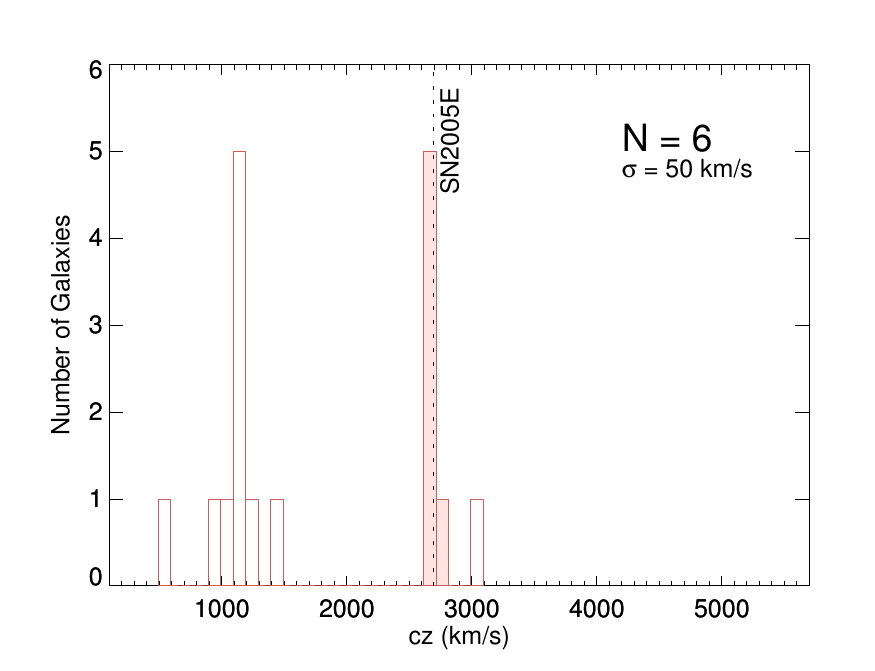} & \includegraphics[width=6cm]{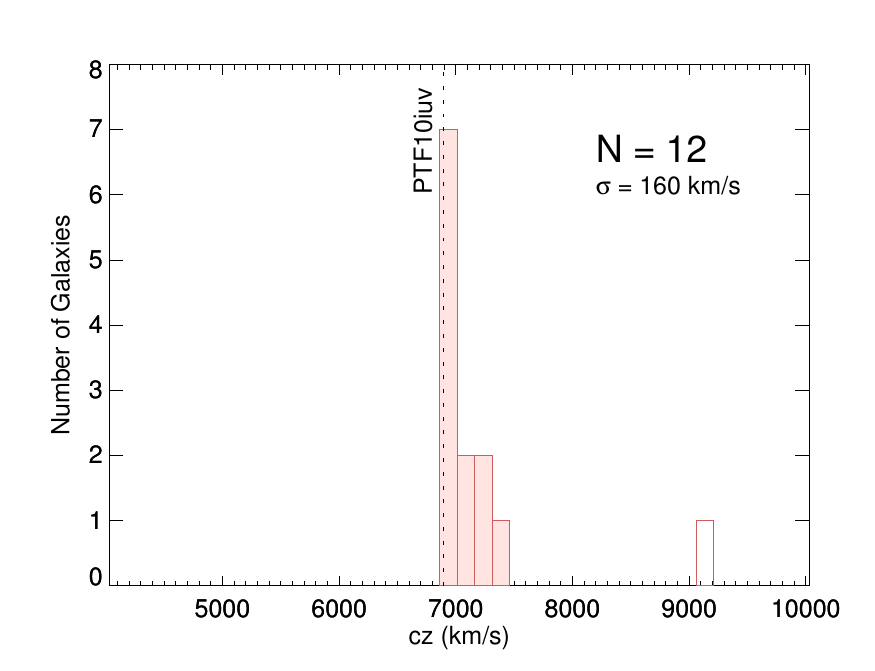} \\
    \includegraphics[width=6cm]{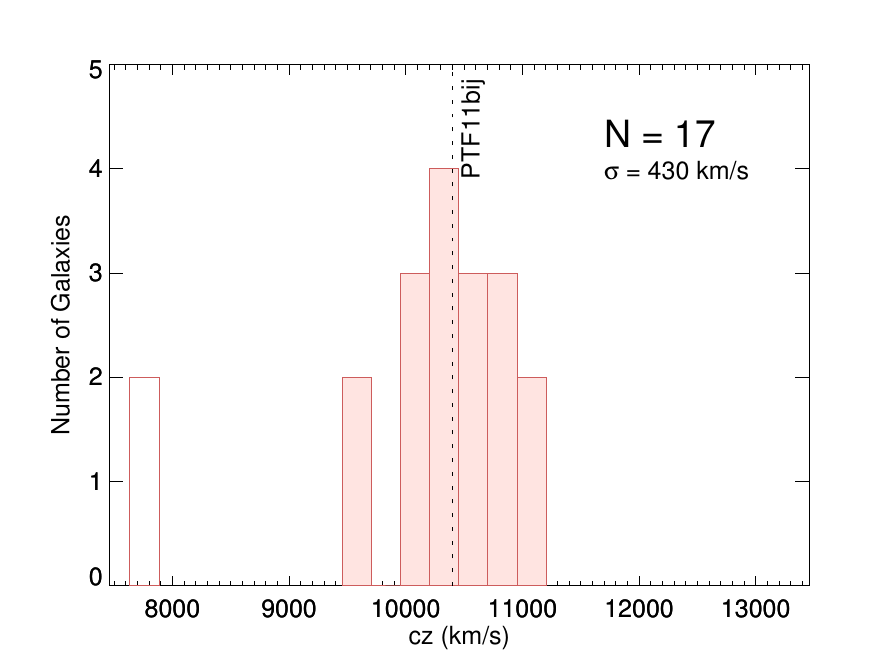} & \includegraphics[width=6cm]{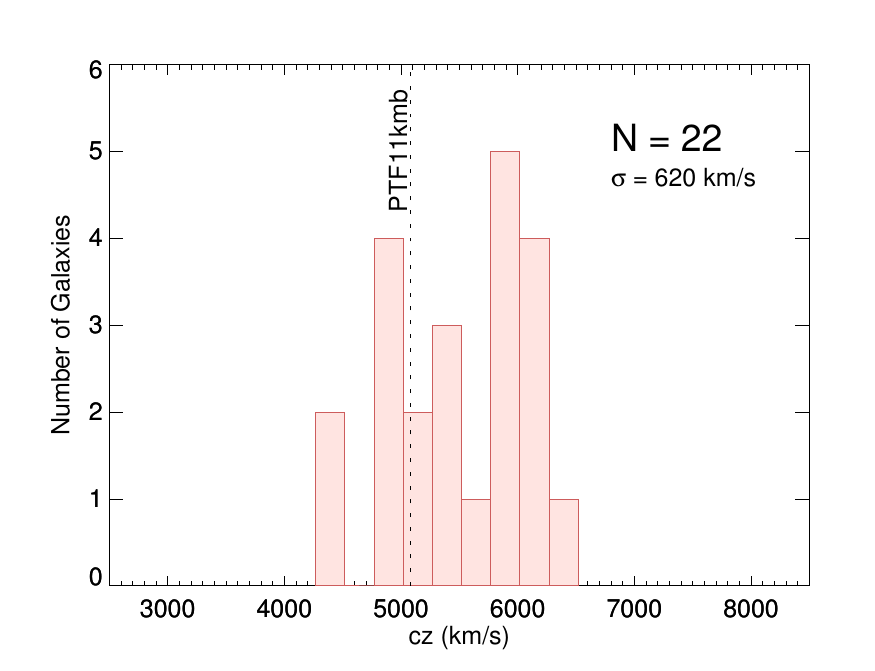} &
    \includegraphics[width=6cm]{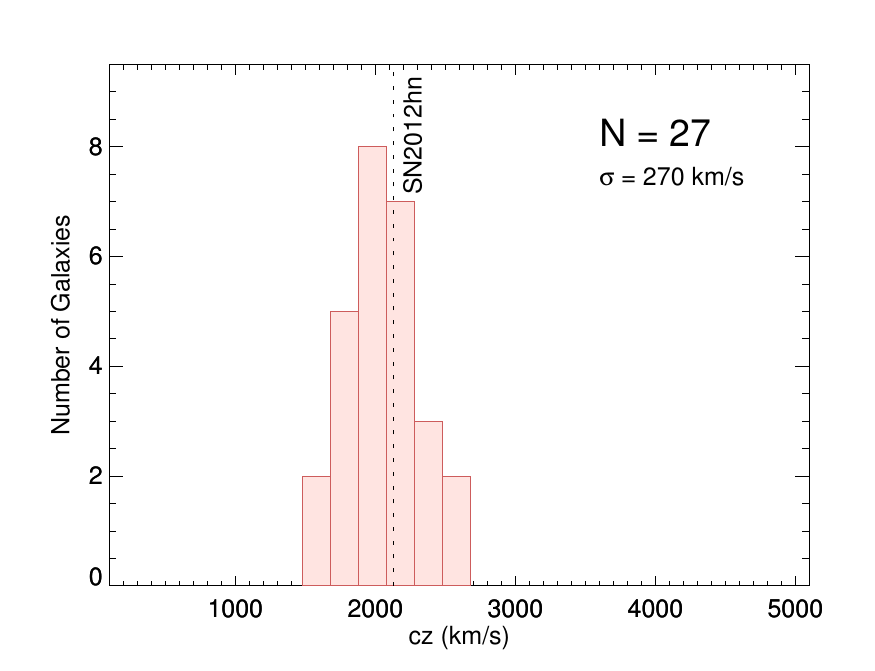} \\
    \includegraphics[width=6cm]{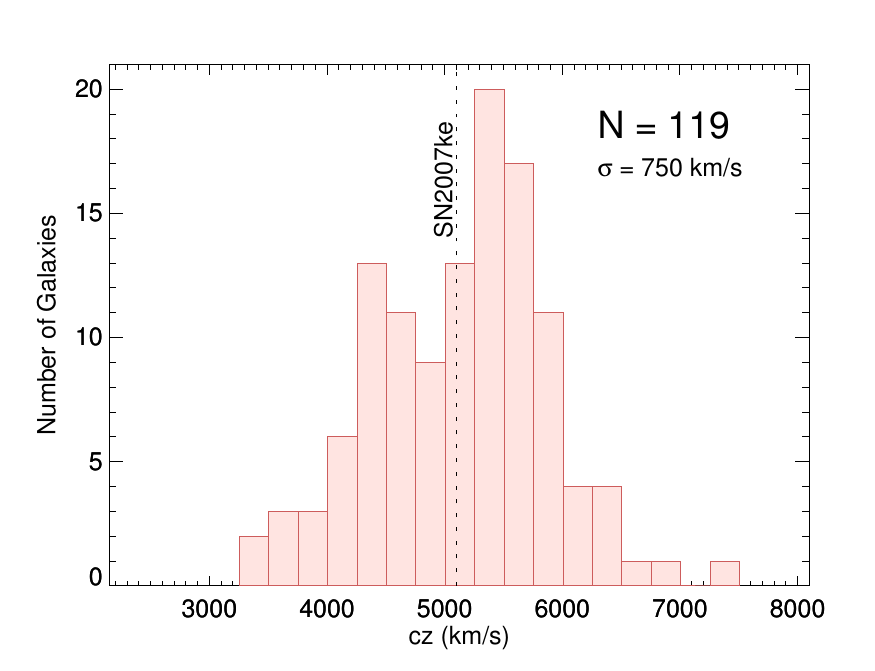} & 
    \includegraphics[width=6cm]{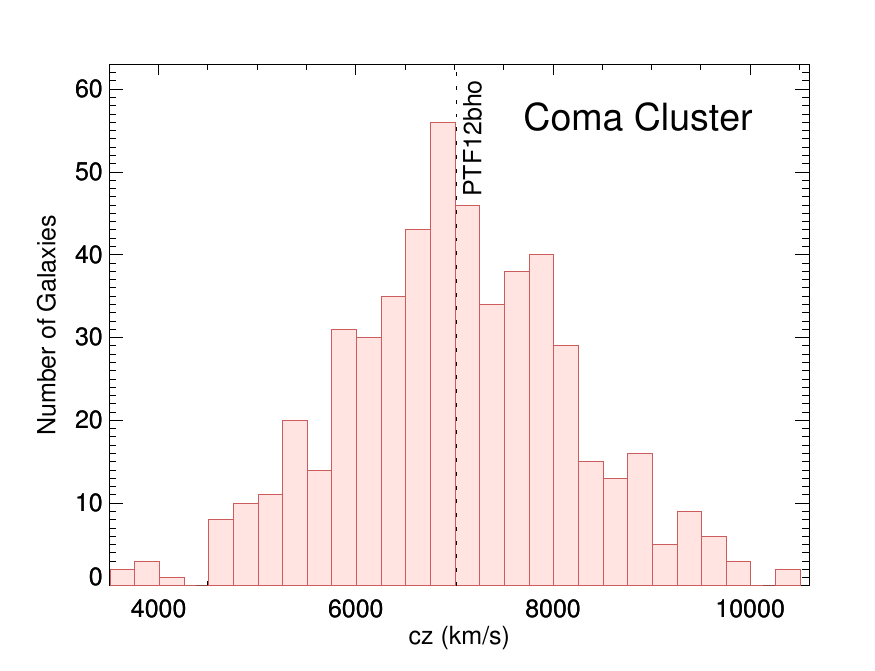}&
    \end{tabular}
    \caption{Radial velocity distributions of galaxies in NED within 1~Mpc radius and $\pm 3000~{\rm km~s}^{-1}$ of the position of each Ca-rich gap transient. The dotted lines mark the redshifts of the (presumed) host galaxy of each transient. The shaded parts of the histograms indicate galaxies that would be group members based on the measured mean velocity and dispersion, while the non-filled part shows other galaxies along the line of sight. The number of galaxies in each group and the line-of-sight velocity dispersion is indicated on each panel. Only PTF09dav appears to come from an isolated galaxy; the remainder are in groups of various richness including two galaxy clusters (SN\,2007ke and PTF12bho). }
    \label{fig:vrad}
\end{figure*}

In addition to their large offsets, it is interesting that both PTF11kmb and PTF12bho are found in group/cluster environment --- both are intra-group or intra-cluster transients. To quantify the extent to which this is also true for other Ca-rich gap transients, we search NED for all galaxies within 1~Mpc and $\pm 3000~{\rm km~s}^{-1}$ of each Ca-rich gap transient. Group members (if any) are then determined by an iterative $3\sigma$-clipping algorithm \citep{yv77,zm98}. We use the biweight estimators for location (central velocity) and scale (velocity dispersion) at each step \citep{bfg90}. The resulting velocity histograms are shown in Figure~\ref{fig:vrad}.

Out of the 8 Ca-rich gap transients confirmed to date, only PTF09dav appears to be in a completely isolated environment, at least to the depth of galaxies with redshifts listed in NED. While NGC\,1032, the host of SN\,2005E, was described as an isolated galaxy by \citet{pgm+10}, there are 5 companions within $\pm 150~{\rm km~s}^{-1}$ in a 1~Mpc radius, making it part of a (sparse) galaxy group. PTF10iuv, PTF11bij, PTF11kmb, and SN\,2012hn are all found in galaxy groups with 10--30 members listed in NED. SN\,2007ke, like PTF12bho, is found in a galaxy cluster. We also note that the Ca-rich gap transients are generally found in galaxies near the center of the velocity distribution, and are in several cases also found near the brightest group or cluster galaxy.

Cluster environments are rare; thus, finding a significant fraction of Ca-rich gap transients there is unlikely to be a coincidence. We can use our measured line-of-sight velocity dispersions to estimate the underlying halo masses following Equation~6 in \citet{ymv+07},
\begin{equation}
\sigma = 397.9~{\rm km~s}^{-1} \left(\frac{M_h}{10^{14}~h^{-1}~{\rm M}_{\odot}}\right)^{0.3214} ,
\end{equation}

\noindent suggesting that about half of our systems have halo masses $\gtrsim 10^{14}~{\rm M}_{\odot}~h^{-1}$. At these masses, the halo mass function starts turning over exponentially (e.g., \citealt{ps74,tkk+08}). While the numbers are low, we conclude that Ca-rich gap transients seem to show a preference for massive/dense environments.

\section{Discussion \& Conclusions}
\label{sec:conc}
We have shown that PTF11kmb and PTF12bho are members of the class of Ca-rich gap transients as described by K12, having low peak luminosities, rapid evolution, and a rapid transition to the nebular phase with a nebular spectrum dominated by Ca features. The addition of these two objects brings the number of confirmed members (with well-sampled photometric and spectroscopic information) to eight. PTF11kmb and PTF12bho are also the Ca-rich gap transients found at the largest offsets from their presumed host galaxies to date, and are perhaps better described as intra-group and intra-cluster transients (e.g., \citealt{gmg+03,sgb+11}).

Taken together, the host environments of the Ca-rich gap transients are striking.  The fact that any are found in early-type galaxies indicates that the progenitors are likely old. We note, however, that the observed fraction of early-type galaxies (7/8) is much higher than what is seen for SNe~Ia and sGRBs, whose host populations include $> 50\%$ late-time galaxies, indicating that their rates depend on both the star-formation rate and stellar mass \citep{slp+06,fbc+13}. Moreover, if the rate of Ca-rich gap transients depended only on stellar mass, we would expect their host-galaxy demographics to follow the distribution of stellar mass in the local universe, and thus find about equal numbers of late-type and early-type hosts \citep{kpf+01,bmk+03}. The fact that the overwhelming majority of hosts are early-type therefore points to other factors being important for the progenitor population.

Given the preference for early-type galaxies, as well as for group and cluster environments, a leading candidate for the dominant factor is stellar population age. In general, massive halos like the ones that host clusters and massive groups formed the majority of their stars before $z = 2$ (e.g., \citealt{bwc13}), and thus are dominated by stellar populations older than $> 10~{\rm Gyr}$. If this is the driving factor for the host-galaxy preference, it implies that the progenitor systems for Ca-rich gap transients have a very long merging time (or more generally, a delay-time distribution that peaks at long timescales). These massive environments are also where the greatest proportion of stars are found in the intra-cluster light (ICL), which may at least partially explain the remote environments \citep{bwc13,gzz05,mhf+05,mhr+13}. If age is indeed the driving factor, however, we would expect to also find a fraction \textit{within} elliptical or S0 galaxies. We note that there are at least two such examples among the candidate Ca-rich gap transients: SN\,2005cz in the elliptical galaxy NGC\,4589, and SN\,2000ds in NGC\,2768 \citep{kmn+10,pgc+11,fcs+03}.

 Our search of the PTF database demonstrates that the relative lack of Ca-rich gap transients at small offsets found in PTF is not likely to be caused by a selection effect, and rather illustrates the importance of wide-field, galaxy-untargeted surveys in detecting such remote transients. One proposed explanation for the offset distribution has been that Ca-rich gap transients originate in very faint systems such as globular clusters, in which case a preference toward group and cluster environments is also predicted \citep{yks+13,smk+15}. However, the deep photometric limits at the locations of Ca-rich gap transients, including those presented in this paper for SN\,2005E and PTF12bho, rule out an underlying dwarf galaxy as well as most of the globular cluster luminosity function. While our \textit{HST} imaging does not rule out a globular cluster origin for PTF11kmb, it is unlikely to be a dominant channel for producing Ca-rich gap transients, at least assuming they are not kicked out of the globular cluster before merging. 

The general lack of underlying/faint host systems is consistent with previous findings \citep{llc+14,llj+16}, and has been used to argue that Ca-rich gap transients do not form ``in situ'' at these remote locations, but have traveled to their explosion sites. One attractive property of the NS-WD merger scenario \citep{met12,smk+15,mm16} is that the binary would have received a kick in the SN explosion that creates the neutron star; such SN kicks are indeed also evoked to explain the offset distribution of sGRBs, which is consistent with theoretical predictions for NS-NS mergers \citep{fb13}. In this context, it is interesting to note that the offset distribution for Ca-rich gap transients is significantly more extreme than that of sGRBs (Section~\ref{sec:offsets}; Figure~\ref{fig:offsets}). In addition, stellar population synthesis models predict lower median velocities for WD-NS systems than for NS-NS systems \citep{brf14}. Thus, if the origin of the large offsets is SN kicks, the merger times for Ca-rich gap transients must be significantly longer than for sGRBs. 

The offset-based argument only holds if we assume that the Ca-rich gap transients originate from the galaxies and thus need to travel a significant distance before they explode. We note, however, that if the progenitor systems are formed in globular clusters, an SN kick could easily eject the system, since the escape velocities of globular clusters are only of order tens of km~s$^{-1}$ (e.g., \citealt{gzp+02}). Thus, if the progenitors are NS-WD binaries, the lack of underlying host systems does not necessarily rule out a contribution from systems originating in globular clusters. In fact, simulations of globular clusters show that in order to match the rate of observed low-mass X-ray binaries, they must assume that a fraction of neutron stars form in electron-capture SNe since most of the neutron star systems forming in core-collapse SNe end up being ejected from the cluster \citep{ihr+08}. A caveat is that only binaries formed on timescales shorter than the time to the SN creating the neutron star would be ejected this way. For binaries formed dynamically after the neutron star is created, one would still expect them to explode within the globular cluster.

If the progenitor system is not a NS-WD binary but rather a WD-WD binary (e.g., \citealt{pgm+10,wsl+11,sfk+12,dh15}), the remote locations require a different origin than SN kicks. One possibility, as noted above, is that group and cluster environments have a large fraction of stars in ICL, resulting from galaxy-galaxy stripping and galaxy mergers in the formation of the cluster. The case for a globular cluster origin, on the other hand, is weaker without being able to invoke SN kicks, but in principle a WD-WD binary could be ejected from a globular cluster through dynamical interactions. A third possibility was proposed by \citet{fol15}, who suggested that Ca-rich gap transients originate in nuclear regions of their host galaxies and are flung out at high velocities following an interaction with the central supermassive black hole. At the very least, any progenitor model will have to be able to explain not just the properties of the Ca-rich transients themselves but also their strong environment preferences.

Finally, we note that the Ca-rich gap transients constitute one emerging subclass amongst a number of peculiar transients that have been discovered over the past decade and that are some combination of faint, fast, hydrogen-poor, or found in old environments and at large offsets (e.g., \citealt{fnc+10,fcc+13,mst+11,kkg+10,isw+15,ssf+13}). While the properties described by K12 were modeled on SN\,2005E and attempted to define the maximum set of characteristics that give the largest number of related transients, they were still empirically defined, so it is possible that they do not all constitute the same physical phenomenon. Notably, K12 allowed for diversity in the photospheric-phase spectra and did not require the detection of He, without which PTF09dav would not have been considered a member of this class. The fact that PTF09dav seems to be an outlier both in terms of its SN and host-galaxy properties suggests that it may have a distinct physical origin from the rest of the objects in the sample. Similarly, while we argue here that PTF12bho should be considered a Ca-rich gap transient at least as described in K12, its photospheric spectrum is unique amongst the objects discovered to date. While sample sizes are currently small, upcoming surveys like the Zwicky Transient Facility and the Large Synoptic Survey Telescope should continue to discover more of these rare transients, shedding light on their true origins.

\acknowledgements
R.L. thanks Andrew Wetzler, Wen-fai Fong, Mark Sullivan, Dan Milisavljevic, Giorgos Leloudas, Jesper Sollerman, and Ryan Chornock for useful discussions, and acknowledges helpful interactions with Lars Bildsten, Eliot Quataert, and Dan Kasen at a PTF Theory Network retreat funded by the Gordon and Betty Moore Foundation through Grant GBMF5076.
We thank J.~Silverman, B.~Dilday, J.~Bloom, B.~Sesar, D.~Levitan, P.~Groot, D.~Perley, A.~Horesh, K.~Mooley, and D.~Xu for assisting with the observations presented in this paper.
The Intermediate Palomar Transient Factory project is a scientific collaboration among the California Institute of Technology, Los Alamos National Laboratory, the University of Wisconsin--Milwaukee, the Oskar Klein Center, the Weizmann Institute of Science, the TANGO Program of the University System of Taiwan, and the Kavli Institute for the Physics and Mathematics of the Universe. 
Support for {\it HST} Program GO-13864 was provided by NASA
through a grant from the Space Telescope Science Institute,
which is operated by the AURA, Inc., under NASA contract NAS 5-26555. We thank F.~Yuan, M.~Sullivan, D.~Perley, R.~M.~Quimby, and S.~B.~Cenko for their contributions to the {\it HST} proposal.
This work was supported by the GROWTH project funded by the National Science Foundation under Grant 1545949.
The National Energy Research Scientific Computing Center, which is supported by the Office of Science of the U.S. Department of Energy under Contract No. DE-AC02-05CH11231, provided staff, computational resources, and data storage for this project.
A.G.-Y. is supported by the EU/FP7 via ERC grant No.
307260, the Quantum Universe I-Core program by the Israeli
Committee for planning and funding, and the ISF,
Minerva and ISF grants, WIS-UK ``making connections,''
and Kimmel and YeS awards.  
A.V.F. is grateful for financial support from
the Christopher R. Redlich Fund, the TABASGO Foundation, and NSF           
grant AST--1211916. 
D.A.H. and C.M. are supported by NSF grant AST--313484. 
This research has made use of the NASA/IPAC Extragalactic Database (NED), which is operated by the Jet Propulsion Laboratory, California Institute of Technology, under contract with the National Aeronautics and Space Administration.
Some of the data presented herein were obtained at the W.M. Keck Observatory, which is operated as a scientific partnership among the California Institute of Technology, the University of California and the National Aeronautics and Space Administration. The Observatory was made possible by the generous financial support of the W.M. Keck Foundation. The authors wish to recognize and acknowledge the very significant cultural role and reverence that the summit of Mauna Kea has always had within the indigenous Hawaiian community.  We are most fortunate to have the opportunity to conduct observations from this mountain.

\textit{Facilities:} \facility{PO:1.2m}, \facility{PO:1.5m}, \facility{PO:Hale}, \facility{Keck:I}, \facility{Keck:II}, \facility{Subaru}, \facility{HST}.

\clearpage
\newpage

\begin{deluxetable}{lccccc}
\tablewidth{0pt}
\tabletypesize{\scriptsize}
\tablecaption{Photometry of PTF11kmb and PTF12bho}
\tablehead{
\colhead{Object} & 
\colhead{Observation Date} & 
\colhead{Phase} &
\colhead{Filter} &
\colhead{Magnitude\tablenotemark{a}} &
\colhead{Telescope} \\
\colhead{} &
\colhead{(MJD)}  & 
\colhead{(rest-frame days)} &
\colhead{} &
\colhead{(AB mag)} &
\colhead{}
}
\startdata
PTF11kmb & 55802.2 & $1.5$ & $B$ & 19.84 $\pm$ 0.09 & P60 \\
PTF11kmb & 55805.2 & $4.5$ & $B$ & 20.00 $\pm$ 0.09 & P60 \\
PTF11kmb & 55805.2 & $4.5$ & $B$ & 20.11 $\pm$ 0.10 & P60 \\
PTF11kmb & 55808.2 & $7.5$ & $B$ & 20.67 $\pm$ 0.14 & P60 \\
PTF11kmb & 55803.2 & $2.6$ & $g$ & 19.47 $\pm$ 0.03 & P60 \\
PTF11kmb & 55806.2 & $5.5$ & $g$ & 19.73 $\pm$ 0.05 & P60 \\
PTF11kmb & 55807.2 & $6.5$ & $g$ & 19.84 $\pm$ 0.06 & P60 \\
PTF11kmb & 55808.2 & $7.5$ & $g$ & 20.18 $\pm$ 0.09 & P60 \\
PTF11kmb & 55809.4 & $8.6$ & $g$ & 20.14 $\pm$ 0.09 & P60  
\enddata
\tablenotetext{a}{Corrected for foreground extinction according to \citet{sf11}.}
\tablecomments{Complete table is available in online journal in electronic form.}
\label{tab:lc}
\end{deluxetable}

\begin{deluxetable}{lccccccc}
\tablewidth{0pt}
\tabletypesize{\scriptsize}
\tablecaption{Summary of Spectroscopic Observations}
\tablehead{
\colhead{Object} & 
\colhead{Observation Date} & 
\colhead{Phase} &
\colhead{Telescope+Instrument} &
\colhead{Grating\tablenotemark{a}} &
\colhead{Filter} & 
\colhead{Exp. time\tablenotemark{a}} &
\colhead{Airmass} \\
\colhead{} &
\colhead{(YYYY MM DD.D)}  & 
\colhead{(rest-frame days)} &
\colhead{} &
\colhead{} &
\colhead{} &
\colhead{(s)} &
\colhead{} 
}
\startdata
PTF11kmb & 2011 Aug. 28.5 & $+0.9$ & KeckI+LRIS & 600/4000,400/8500 & none & 450,450 & 1.17 \\
PTF11kmb & 2011 Sep. 21.3 & $+24.2$ & P200+DBSP & 600/4000,158/7500 & none & 600,600 & 1.07 \\
PTF11kmb\tablenotemark{b} & 2011 Oct. 31.4 & $+63.6$ & KeckII+Deimos & 600ZD & GG455 & 2400 & 1.19 \\
PTF11kmb & 2011 Nov. 26.4 & $+89.2$ & KeckI+LRIS & 400/3400,400/8500 & none & 1200,1180 & 1.95 \\
PTF11kmb & 2011 Dec. 31.2 & $+123.5$ & KeckI+LRIS & 400/3400,400/8500 & none & 3600,3480 & 1.49 \\
PTF12bho & 2012 Mar. 15.6 & $+9.1$ & KeckI+LRIS & 600/4000,400/8500 & none & 300,300 & 1.15 \\
PTF12bho & 2012 Mar. 20.5 & $+13.9$ & KeckII+Deimos & 600ZD & GG455 & 600 & 1.03 \\
PTF12bho & 2012 Mar. 23.4 & $+16.7$ & KeckI+LRIS & 400/3400,400/8500 & none & 900,870 & 1.28 \\
PTF12bho & 2012 Apr. 27.4 & $+50.9$ & KeckI+LRIS & 400/3400,400/8500 & none & 3600,3360 & 1.02 \\
PTF12bho & 2012 July 16.3 & $+129.0$ & KeckII+Deimos & 600ZD & GG455 & 3600 & 1.44 \\
PTF11kmb mask & 2014 July 02.5 & \nodata & KeckII+Deimos & 600ZD & GG455 & 2400 & 1.07 
\enddata
\tablenotetext{a}{Comma-separated values indicate setup for blue and red arms, respectively.}
\tablenotetext{b}{As part of a slitmask, aiming to get redshifts of potential host galaxies.}
\label{tab:spec}
\end{deluxetable}

\begin{deluxetable}{lcccc}
\tablewidth{0pt}
\tabletypesize{\scriptsize}
\tablecaption{Summary of \textit{HST} Observations}
\tablehead{
\colhead{Object} & 
\colhead{Observation Date} & 
\colhead{Instrument} &
\colhead{Filter} &
\colhead{Total Exp. Time} \\
\colhead{} &
\colhead{(YYYY MM DD)}  & 
\colhead{} &
\colhead{} &
\colhead{(s)} 
}
\startdata
 PTF11kmb  & 2015 July 12 & WFC3/UVIS & F606W & 5469  \\
 SN\,2005E & 2014 Dec. 10 & WFC3/UVIS & F606W & 5361 
\enddata
\label{tab:hst}
\end{deluxetable}

\begin{deluxetable}{lccccc}
\tablewidth{0pt}
\tabletypesize{\scriptsize}
\tablecaption{Redshifts of Galaxies near PTF11kmb}
\tablehead{
\colhead{Galaxy ID} &
\colhead{$\alpha$(J2000)} &
\colhead{$\delta$(J2000)} &
\colhead{Redshift} &
\colhead{In group?} &
\colhead{Telescope/Instrument}
}
\startdata
2MASX J22223308+3616504 & \ra{22}{22}{33.08} & \dec{+36}{16}{50.4} & 0.0179 & Yes & KeckII/Deimos \\
GALEXASC J222249.41+361811.7 & \ra{22}{22}{49.41}& \dec{+36}{18}{11.7} & 0.02099 & Yes & P200/DBSP \\
GALEXASC J222226.82+362011.9 & \ra{22}{22}{26.82} & \dec{+36}{20}{11.9} & 0.166 & No &  KeckII/Deimos \\
(anon.)  & \ra{22}{22}{52.30} & \dec{+36}{17}{39.9} & 0.166 & No & KeckII/Deimos \\
(anon.)  & \ra{22}{22}{37.03} & \dec{+36}{20}{43.6}& 0.166 & No & KeckII/Deimos \\
(anon.)  & \ra{22}{22}{42.66} & \dec{+36}{19}{25.3} & 0.166 & No & KeckII/Deimos \\
(anon.)  & \ra{22}{23}{05.50} & \dec{+36}{19}{00.5} & 0.403 & No & KeckII/Deimos 
\enddata
\label{tab:11kmbgal}
\end{deluxetable}

\appendix
\section{Ca-Rich Gap Transient Candidate PTF10hcw}
\label{sec:10hcw}

\begin{figure}
\centering
\includegraphics[width=3.5in]{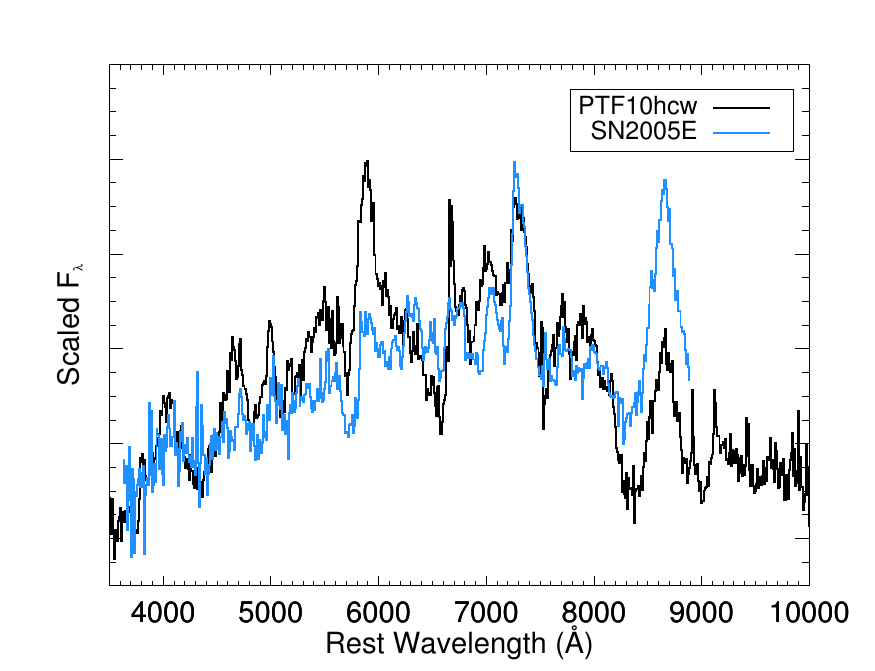}
\caption{Our single spectrum of PTF10hcw (black), compared to SN\,2005E (blue). The PTF10hcw spectrum is taken at phase 13~days past discovery (and 18~days after the last nondetection), while the SN\,2005E spectrum is at phase 24~days past discovery. Like SN\,2005E, PTF10hcw developed [\ion{Ca}{2}] $\lambda\lambda$7291, 7324 emission at an early phase. }
\label{fig:10hcw_spec}
\end{figure}

PTF10hcw was discovered in P48 data at (J2000) $\alpha =$ \ra{08}{43}{36.22}, $\delta =$ \dec{+50}{12}{38.5} on 2010 May 02.2, at a discovery magnitude of $R = 19.3~{\rm mag}$. No emission was detected at this location to a limit $R > 20.3~{\rm mag}$ on 2010 Apr. 27. The object was classified as a SN~Ib based on an LRIS spectrum taken on 2010 May 15 \citep{gaa+10}, which is shown in Figure~\ref{fig:10hcw_spec}. 

\begin{figure}
\centering
\includegraphics[width=3.5in]{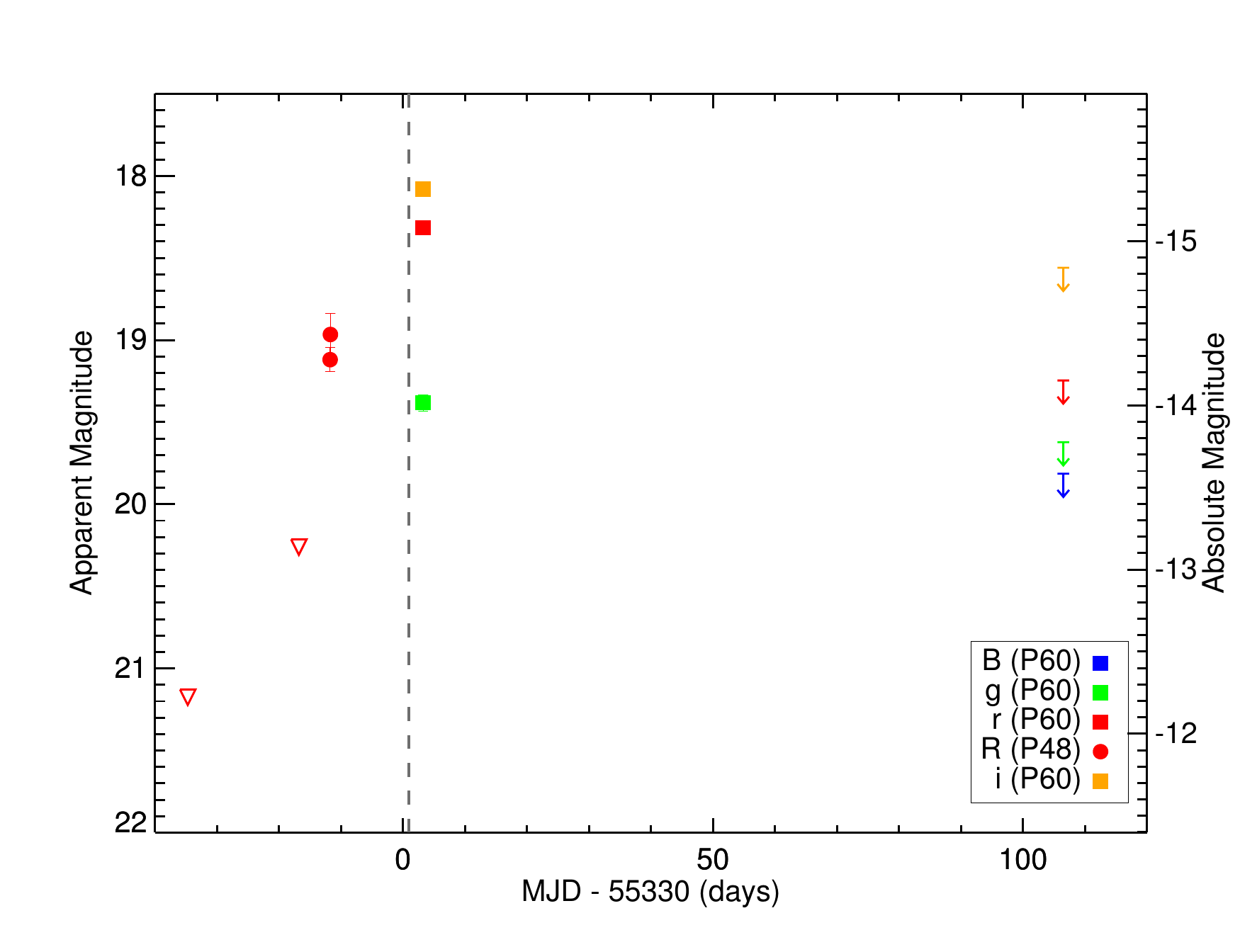}
\caption{Light curve of PTF10hcw. Triangles and arrows show upper limits from P48 and P60, respectively. The light curve is extremely sparse as the field went into solar conjunction shortly after the transient was discovered, hence the 100-day gap. Neither the rise time nor the peak magnitude are very well constrained, but based on the available data it is likely that the peak magnitude is comparable to that of the Ca-rich gap transients (Figure~\ref{fig:lccomp}). The time of the spectrum displayed in Figure~\ref{fig:10hcw_spec} is shown by the dashed line. }
\label{fig:10hcw_lc}
\end{figure}

We suggest that PTF10hcw may have been a Ca-rich gap transient based on the following two observations. First, the spectrum shows emerging [\ion{Ca}{2}] emission, which is not present in normal SNe~Ib at such an early phase (e.g., \citealt{lmb+16}), but was also seen in SN\,2005E. Second, while the light curve (shown in Figure~\ref{fig:10hcw_lc}) for this object is extremely sparse since it was discovered as the field was going into solar conjunction, the brightest observed $r$-band magnitude was $M_r = -15.1~{\rm mag}$. This was at a phase 15~days after discovery (and 21~days after the last nondetection), so it is likely that the SN was near peak light, and by extension that the peak brightness was fainter than that of a typical SN~Ib. This is, in itself, not enough to classify PTF10hcw as a Ca-rich gap transient as opposed to a peculiar Type Ib SN, as we lack both a well-sampled light curve and a nebular-phase spectrum; we merely present it here as a candidate.

\begin{figure}
\centering
\includegraphics[width=3in,angle=270]{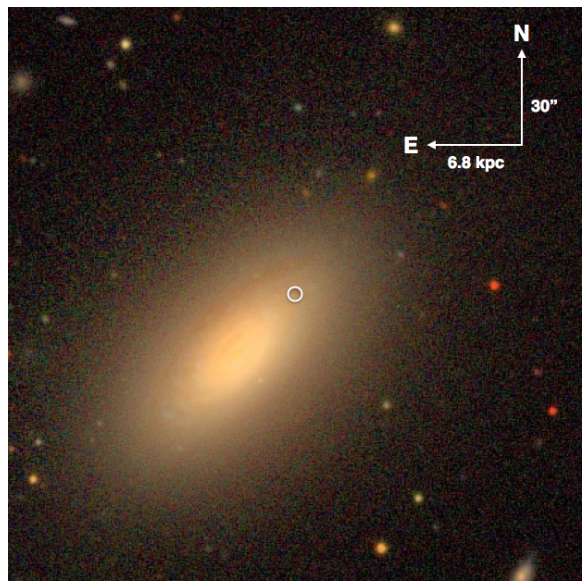}
\caption{3\arcmin $\times$ 3\arcmin\, composite $gri$ SDSS image of NGC\,2639, the host galaxy of PTF10hcw. The image is centered on the location of the transient, which is shown by the white circle.}
\label{fig:10hcw_img}
\end{figure}

We note that PTF10hcw was found at a projected offset 25.7\arcsec (5.9~kpc) from the nucleus of NGC\,2639, which is classified as an Sa galaxy with LINER nuclear activity in NED. The host galaxy and SN location are shown in Figure~\ref{fig:10hcw_img}. NGC\,2639 is the brightest galaxy in a group of 7 members with redshifts in NED, and with a line-of-sight velocity dispersion of 130~km~s$^{-1}$.

\end{document}